\documentclass[amsmath,amssymb,aps,onecolumn,prb,longbibliography,floatfix,superscriptaddress]{revtex4-2}
\usepackage{graphicx,bm,times}
\usepackage[utf8]{inputenc}
\usepackage{braket}
\usepackage{amsfonts}
\usepackage{color}

\renewcommand{\Re}{\operatorname{Re}}

\usepackage[breaklinks]{hyperref}
\hypersetup{
    pdfstartview={FitH},
    colorlinks=true,
    linkcolor=blue,
    citecolor=blue,
    urlcolor=blue
}

\begin{document}

\title{Electron momentum densities from QSGW and $G^0W^0$: Revealing the role of many-body effects within the reduced density matrix}

\author{A.~D.~N.~James}
\email{aj12959@bristol.ac.uk}

\affiliation{H.~H.~Wills Physics Laboratory,
University of Bristol, Tyndall Avenue, Bristol, BS8 1TL, United Kingdom}

\author{J.~A.~Gould}
\affiliation{H.~H.~Wills Physics Laboratory,
University of Bristol, Tyndall Avenue, Bristol, BS8 1TL, United Kingdom}

\author{T.~M.~Mason}
\affiliation{H.~H.~Wills Physics Laboratory,
University of Bristol, Tyndall Avenue, Bristol, BS8 1TL, United Kingdom}

\author{J.~Jackson}
\affiliation{Scientific Computing Department, 
STFC Daresbury Laboratory, Warrington WA4 4AD, United Kingdom}

\author{S.~B.~Dugdale}
\affiliation{H.~H.~Wills Physics Laboratory,
University of Bristol, Tyndall Avenue, Bristol, BS8 1TL, United Kingdom}

\date{\today}

\begin{abstract}
The ground-state many-body electron momentum density, which can be probed by x-ray Compton scattering, holds insights into the electronic structure of materials. Comparisons between the measured so-called Compton profiles and the theoretical ones are invaluable in assessing the successes and failures of the methodology used to generate the theoretical ground-state electronic structure. 
Here, we present calculations of the Compton profiles of Li, Si, Cr, and Ni using the state-of-the-art QSGW method within the Questaal package compared with density functional theory (DFT), one-shot $GW$ ($G^0W^0$) predictions and with experiment.
This comparison reveals significant differences between the QSGW and $G^0W^0$ methods which we attribute to the distinction between the single particle density provided by the QSGW method and the many-body density that we construct from the $G^0W^0$ theory; although in general the QSGW description of the electronic structure is superior to that of $G^0W^0$, we find the use of the many-body reduced density matrix is key to improving the agreement of the Compton profile with experiment.
\end{abstract} 

\maketitle

\section{Introduction}

The electron momentum density (EMD) and its spin dependence, as probed by Compton scattering and magnetic Compton scattering, provide direct insight into the ground-state many-body wavefunction~\cite{cooper_x-ray_2004,dugdale:14,cooper_spin_2000,barbiellini:13}. 
Comparisons of the EMD calculated using density functional theory (DFT) with experiment can show significant differences which have been largely assigned to the application of the Kohn-Sham formalism, where the approximate treatment of electronic correlation effects causes error in the ground-state density.
Compton scattering can probe the electron correlations within complex materials where such effects are more pronounced and more challenging to describe theoretically~\cite{ruotsalainen2018isotropic,billington2015magnetic,James_2023}. However, even simple elemental materials such as Li~\cite{sakurai:95,schulke:01}, Si~\cite{Erba_2011,Pisani_2011}, Cr~\cite{DUGDALE2000361}, and Ni~\cite{di.du.98}, show significant discrepancies between the measured Compton profiles and their DFT calculations, contrary to the expected accurate descriptions from the DFT Kohn-Sham framework with either the local density approximation (LDA) or generalised gradient approximation (GGA). 
Indeed, recent studies beyond-DFT methods~\cite{mandal2022electronic,Friedrich-QSGW} have illustrated that significant correlation effects must be acknowledged even within alkali and alkaline-earth metals.
More sophisticated theoretical methods, which can account for electron correlations more accurately than is possible using the available DFT models, are necessary to allow us to better understand the modification to the ground-state density due to electronic correlation.

One common beyond-DFT method is Hedin's $GW$ method~\cite{hedin_new_1965}. The $GW$ approximation provides an accurate method for describing the electronic structure of solids that addresses the most significant failings of DFT without {\it ad hoc} parameters or corrections.
GW is typically applied as a perturbation to a simpler description of the electronic structure, such as DFT, and the quality of the results depends upon the quality of the starting point.
Quasiparticle self-consistency (QSGW) improves upon $GW$ by providing a starting-point-independent single particle description by building a sequence of perturbative $GW$ calculations, each starting from an effective potential derived from the previous $GW$ calculation, iterating until self-consistency is achieved.
The QSGW prescription has the property that its non-interacting (single particle) bands match the energetic positions of the poles in the spectral function of the interacting $GW$ theory (when evaluated using the QSGW single particle states).

The central question addressed in this work is whether improvements in quasiparticle electronic structure, as achieved within QSGW, necessarily translate into improved electron momentum densities and Compton profiles. We use the Questaal package~\cite{questaal-cpc,questaal-website} to calculate the QSGW EMDs. We also verify that the DFT EMDs from Questaal are consistent with those from the established Elk code~\cite{elk,Ernsting_2014}. Since the EMD is determined by the reduced density matrix rather than solely by quasiparticle energies, comparisons between QSGW and $G^0W^0$ provide a direct test of the importance of many-body occupation effects beyond an effective single-particle description. The four materials considered here tackle distinct aspects of this problem: Li tests correlation effects in a simple metal, Si probes bonding and insulating behaviour, Cr probes Fermiology, and Ni probes spin-polarised momentum densities.

\section{Background}

\subsection{QSGW}
The $GW$ method provides a simple, low order approximation for the electronic self-energy, $\Sigma^{GW} = iGW$, in terms of the screened Coulomb interaction $W$~\cite{hedin_new_1965}.
In practical calculations, the interacting Green's function $G$ is commonly approximated by a non-interacting reference Green's function $G^0$, typically constructed from a DFT calculation. The screened Coulomb interaction $W$ is likewise approximated by $W^0$, obtained within the random phase approximation (RPA) using the DFT electronic structure, where the polarisability is given by $\Pi = iG^0G^0$. The resulting self-energy is therefore $\Sigma^{GW} \approx iG^0W^0.$
The accuracy of the resulting $G^0W^0$ calculation depends on the quality of the starting point used to construct $G^0$, including the choice of DFT exchange-correlation functional. In practice, however, $G^0W^0$ calculations based on DFT have been highly successful in improving the electronic structure of many weakly correlated materials, particularly when compared with spectroscopic probes such as angle-resolved photoemission spectroscopy (ARPES) and ellipsometry experiments~\cite{gw_compendium}.
For materials showing stronger electronic localisation (notably transition metal oxides) the $G^0W^0$ approach suffers from the inaccuracy of the starting point.

Quasiparticle self-consistent $GW$~\cite{qsgw-original} is based on constructing an effective single-particle potential $V^{\textrm {QSGW}}$ based on the interacting self-energy,
\begin{equation}
V^{QSGW}=\frac{1}{2}\sum\limits_{ij}|\psi_i\rangle\left(\Re[\Sigma(\varepsilon_i)]_{ij}+\Re[\Sigma(\varepsilon_j)]_{ij}\right)\langle\psi_j|\, .\label{eq:veff-qsgw}
\end{equation}
This new potential is solved, replacing DFT's $V_{xc}$ to provide a new set of single particle eigenvalues and eigenfunctions; using these, the $G^0W^0$ cycle is repeated until convergence is obtained in $V^{QSGW}$.

QSGW shows significantly improved accuracy over density functional methods in a wide range of observables: electronic band gaps and band widths, Fermi surfaces, effective masses, as well as magnetic and dielectric properties are consistently improved~\cite{questaal-cpc}.
This makes QSGW a benchmark method for accurate, parameter-free materials simulation: it includes the Fock exchange exactly, and is therefore ``self-interaction free'', and correctly treats both strongly localised and delocalised electrons on the same footing, without the need of designating subspaces (specific manifolds of bands) for special treatment.
Because $V^{QSGW}$ forms an essentially optimal single-particle starting point for $GW$, whose bands precisely match the peaks of the interacting $GW$ spectral function, it also provides a non-interacting density that provides an optimal effective single-particle representation of the interacting $GW$ quasiparticle spectrum.
While the $GW$ self-energy is explicitly frequency dependent and non-local, the QSGW effective potential $V^{\textrm{QSGW}}$ is non-local but static.

\subsection{Compton scattering}
\label{CP_background}

The EMD, $\rho(\mathbf{p})$, which is most simply defined as the square modulus of the momentum-space ground state many-body wavefunction, can be experimentally accessed through x-ray Compton scattering, the inelastic scattering of photons by electrons~\cite{cooper:85,cooper_compton_1997,cooper_x-ray_2004}.
Within the impulse approximation, the measured double-differential inelastic photon-electron scattering cross-section $\sigma_c$ (with respect to measured X-ray energy $\omega$ and solid angle $\Omega$) is directly proportional to the Compton profile, 
\begin{equation}
\frac{\mathrm{d}^2\sigma_c} {\mathrm{d}\Omega \mathrm{d}\omega} \propto J(p_z)\, ,
\end{equation}
where the Compton profile, $J(p_z)$, is the one-dimensional projection of the EMD (i.e. a double integral of two momentum components perpendicular to the scattering vector, conventionally denoted $p_z$, which is aligned with a chosen crystallographic direction):
\begin{equation}
J(p_{z})= \iint \left[ \rho_\uparrow(\mathbf{p})+\rho_\downarrow(\mathbf{p}) \right]\, \mathrm{d} p_{x} \mathrm{d} p_{y}\, . \label{eq:J}
\end{equation}
In this equation, $\uparrow$ and $\downarrow$ denote spin states and $\rho_\uparrow(\mathbf{p})+\rho_\downarrow(\mathbf{p})$ is the total EMD. 
Since photons can only be scattered from the occupied electron momentum states, Compton scattering is sensitive to the Fermi surfaces of metals \cite{dugdale:14}.
If the incident photon beam has some circular polarisation, the scattering cross-section contains a term  which depends on the spin of the scattering electron~\cite{sakai_magnetic_1996,cooper_spin_2000}. 
The magnetic Compton profile (MCP) $J_{\mathrm{mag}}(p_z)$ is the projection of the spin density on the scattering momentum,
\begin{equation}
 \label{eq:J_mag}
  J_{\mathrm{mag}}(p_z)= \iint 
    \left[
      \rho_\uparrow(\mathbf{p})-\rho_\downarrow(\mathbf{p})
    \right]
    \mathrm{d}p_x \mathrm{d}p_y\, .
\end{equation}

Compton profiles are typically measured along high-symmetry directions; when the full three-dimensional EMD is needed, it can be reconstructed from several (typically 10---20) Compton profile measurements along various crytallographic directions~\cite{kontrym-sznajd_fermiology_2009,ketels_momentum_2021}.

The EMD is itself directly related to the reduced density matrix $n_\sigma(\mathbf{r}, \mathbf{r}')$ ~\cite{olevano_momentum_2012,barbiellini2001EMD}
\begin{equation}
\rho_\sigma(\mathbf{p}) = \int\limits_{V}
\, n_\sigma(\mathbf{r}, \mathbf{r}') e^{-i \mathbf{p} \cdot (\mathbf{r} - \mathbf{r}')}d\mathbf{r}\, d\mathbf{r}',
\end{equation}
and if the reduced density matrix can be diagonalised and thus be expressed as $ n_\sigma(\mathbf{r}, \mathbf{r}')=\sum_{\eta\mathbf{k}}n^{\sigma}_{\eta\mathbf{k}}\psi^\sigma_{\eta\mathbf{k}}(\mathbf{r})\psi^{\sigma*}_{\eta\mathbf{k}}(\mathbf{r}')$, then the EMD can be presented in terms of single-particle states,
\begin{equation}
  \rho_\sigma(\mathbf{p})= \sum_{\eta\mathbf{k}}n^{\sigma}_{\eta\mathbf{k}}\left|\int\limits_{V}  \psi^\sigma_{\eta\mathbf{k}}(\mathbf{r})e^{-i \mathbf{p} \cdot \mathbf{r}} \mathrm{d} \mathbf{r}\right|^{2}\, , \label{EMD}
\end{equation} 
where $\psi^\sigma_{\eta\mathbf{k}}(\mathbf{r})$ are the real-space natural orbitals to be Fourier transformed, $V$ is the volume of the crystal and $n^{\sigma}_{\eta\mathbf{k}}$ are occupation distributions with state index $\eta$, and spin index $\sigma$.
For independent-particle descriptions, such as DFT or QSGW, Eqn.~\ref{EMD} is evaluated using the eigenstates of the corresponding effective single-particle Hamiltonian: the Kohn--Sham orbitals in DFT, and the eigenstates of the static effective QSGW potential in QSGW. In this case the occupation factors $n^\sigma_{\eta\mathbf{k}}$ are those of an independent-particle system. For an interacting many-body system, however, the more fundamental object is the one-body reduced density matrix itself. If this density matrix is diagonalised, Eqn.~\ref{EMD} may be written in the same form using its natural orbitals and their generally fractional occupation numbers~\cite{olevano_momentum_2012,james2021magnetic,James_2023}.

Furthermore, it should be noted that the Kohn-Sham formalism gives the many-body real-space ground-state density by construction, but this is not guaranteed for the EMD due to it being constructed from the Fourier transformed fictitious Kohn-Sham wave functions. Lam and Platzman derived a correction term due to the correlations between the Kohn-Sham wave functions~\cite{lam_momentum_1974_I,lam_momentum_1974}. Typically this correction is an isotropic contribution to the EMD which cannot account for all the dispcrepancies between experiment and theory.

Additional information can be extracted using the Fourier transform of the Compton profile, which is the auto-correlation function denoted as $B(\textbf{r})$~\cite{cooper_x-ray_2004,cooper:85}. The auto-correlation function has additional features which are useful for analysing the experimental data. The value of $B(\textbf{r})$ at the origin, $B(\textbf{r} = 0)$, corresponds to the number of electrons. For each Bravais lattice vector $\textbf{R}$, $B(\textbf{R}) = 0$ is strictly true for the single-particle picture of materials without a Fermi surface. Lastly, owing to the convolution theorem, the convolution of the Compton profile with an experimental resolution function becomes a multiplicative factor within the $B(\textbf{r})$. Therefore $B(\textbf{r})$ is another powerful quantity obtainable from Compton scattering measurements which is used to help understand materials.

\begin{figure*}[!ht]
        \centerline{\includegraphics[width=0.9\linewidth]{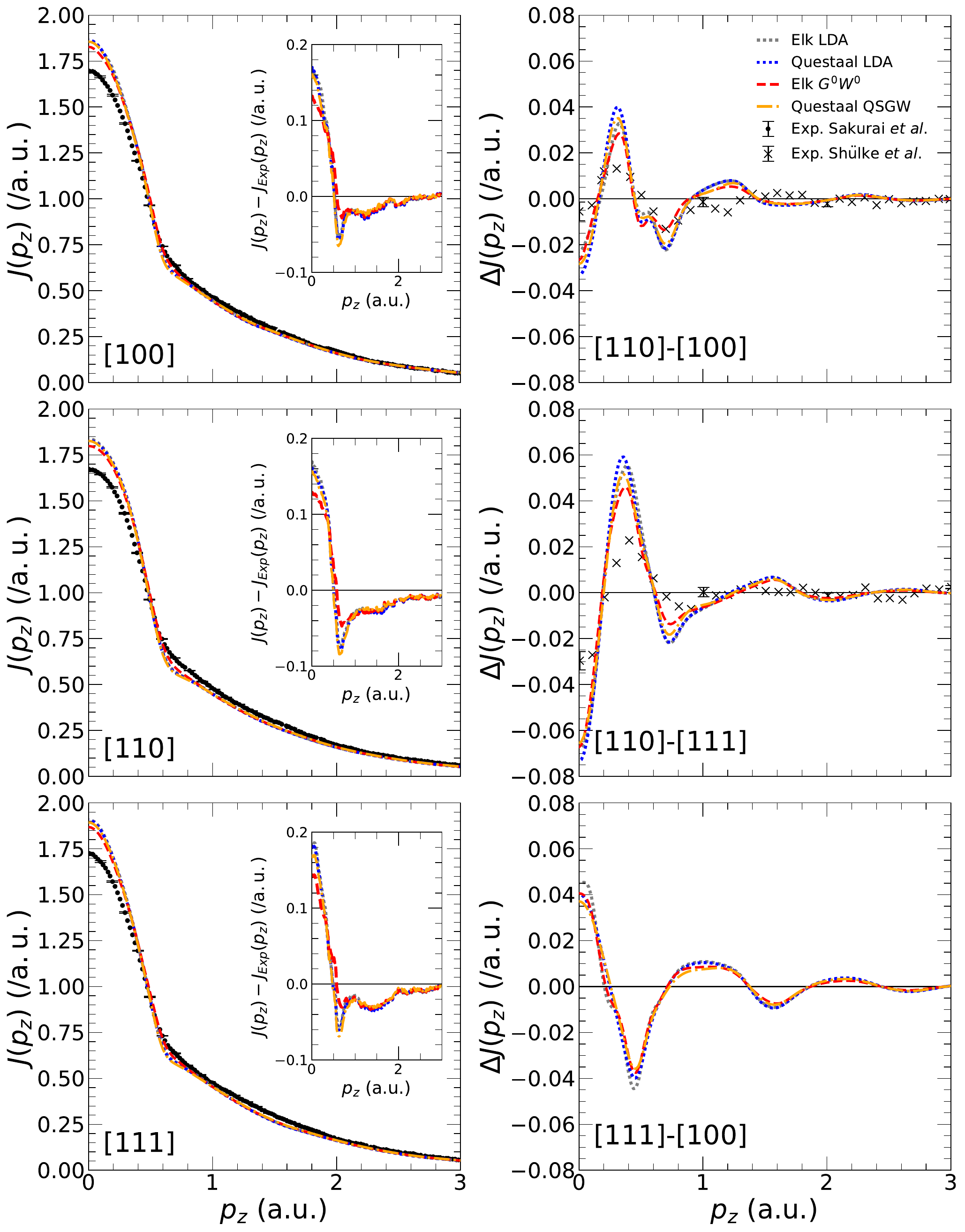}}
        \caption{
        Left column: Comparison of three high-symmetry direction ([001], [011], [111]) experimental Li Compton profiles (CPs) from Sakurai~\textit{et al.} ~\cite{Sakurai-Li} (dots) with our Elk (LDA), Elk $G^0W^0$, Questaal DFT, and Questaal QSGW CPs. 
        To avoid clutter, the error bars are shown for every fifth data point and are estimated from the relative percentage error reported at $p_z=0$ of the Sch\"ulke~\textit{et al.}~\cite{Schulke-Li} and only serve as an indication of the scale of the error. The insets show the differences between the theory and the experimental data (without error bars) along each high symmetry direction. 
        Right column: The theoretical directional differences of the high symmetry directions ([001], [011], [111]) from the left column. The experimental directional differences, with errors, are from Sch\"ulke~\textit{et al.}~\cite{Schulke-Li}. The [111]-[100] experimental directional difference is not available within Ref.~\cite{Schulke-Li} and hence not included. 
        The data is presented up to a momentum of 3 a.u. which is the same as that shown for the experimental CPs.
        We convolute our calculations with an experimental resolution function with respect to the experimental data they are being compared to. The calculations in the left column uses the resolution function with 0.12 a.u. Full-Width-at-Half-Maximum (FWHM) and those in the right column uses the function with 0.14 a.u. FWHM. 
        }
        \label{fig:Li-theory-vs-experiment}
\end{figure*}

\section{Methods}
DFT and QSGW calculations were performed using the Questaal suite~\cite{questaal-cpc} with automatic setup of basis parameters.
Eigenfunctions of the converged $V^{\textrm{QSGW}}$ were expanded in plane waves for dense $k$-meshes (Monkhorst-Pack grids of size $40^3$ points) up to a large momentum cut-off, each function being normalised to unity within this basis.
Questaal's single-particle solver uses an all-electron linear muffin-tin orbital basis which gives access to the full wavefunction without pseudoisation.
Calculations at the DFT level were repeated using the all-electron full-potential linearised augmented-plane wave (with local orbitals) (L)APW+lo Elk code~\cite{elk} to provide an independent verification of the Questaal EMD implementation. 
These Elk calculations were converged on dense Monkhorst-Pack $k$-meshes of $20^3$ points up to a large momentum cut-off. For the exchange-correlation approximation, both Questaal and Elk used the local density approximation (LDA) and the spin version LSDA for magnetic Ni~\cite{perdew1992}. We also present results from the Elk one-shot $G^0W^0$ Matsubara method implementation. 
For our $G^0W^0$ calculations, we started from a converged DFT (LDA/LSDA) result and used a dense $q$-point mesh commensurate with the $k$-point mesh. We obtained a new Green's function $G$ from the $G^0W^0$ self-energy by solving the Dyson equation. The Fermi level was updated when calculating this new Green's function to conserve the charge. We evaluated the $G^0W^0$ self-energy using a sufficiently high Matsubara frequency cut-off to reduce the associated numerical error.

To calculate the valence EMD for each theoretical method, we used a linear tetrahedron method \cite{Ernsting_2014} to evaluate Eqn.~\ref{EMD} and set a maximum momentum cut-off to where we found the valence electron EMD to be approximately zero beyond it. For the $G^0W^0$ EMDs from Elk, we calculated the valence EMD from the reduced density matrix derived from the $G^0W^0$ natural orbitals and occupations~\cite{James_2025,james2021magnetic,olevano_momentum_2012}; it is worth emphasising that this method does not require analytic continuation. For Ni, we used the same Elk parameters as in Ref.~\cite{James_2025} and we present the fixed-spin moment magnetic Compton profiles where the spin moment has been fixed to the experimental spin moment of 0.56$\mu_{\rm B}$. 
We calculate (magnetic) Compton profiles by evaluating Eqn.~\ref{eq:J} or Eqn.~\ref{eq:J_mag} with respect to the specified scattering vector. 
In explicitly stated cases, we convoluted our theoretical Compton profiles with a Gaussian function with a full-width-at-half-maximum (FWHM) equal to that of the experimental resolution. 
For all calculated valence Compton profiles, we normalised their total integral to the appropriate number of valence electrons. For magnetic systems, the normalised Compton profiles are used to calculate the magnetic Compton profiles which then have integrals equal to the spin magnetic moment. Both the presented experimental and theoretical Compton profiles for each material correspond to the same valence electron configuration.  The calculated electron momentum densities, together with all input files, are available in the data archive associated with this paper~\cite{data-archive}.

\section{Results}
\subsection{Li}

Li has been extensively studied by Compton scattering because it provides an excellent opportunity to study the role of electron correlation in a truly simple metal. 
There are several high-resolution Compton scattering measurements that extract the renormalisation factor~\cite{Hiraoka-Li-Z-Compton}, the temperature dependence of the Compton profile~\cite{Sternemann-Li-temp-CPs,Chen-Li-temp-CPs}, Fermi surface anisotropy~\cite{Sakurai-Li,Schulke-Li}, and the full three-dimensional (3D) reconstructed EMD and occupation density~\cite{Schulke-Li}.
Early measurements already revealed a clear discrepancy with DFT predictions, the experimental profiles being broader with a Fermi surface smearing beyond that attributable to instrumental resolution; these discrepancies persist even when the Lam-Platzman correction~\cite{lam_momentum_1974,lam_momentum_1974_I} to the DFT(LDA)~\cite{Sakurai-Li,Schulke-Li} is included.
The principle differences are that the predicted Compton profiles are too large at small momentum and too small around $\mathbf{k}_F$ and beyond.
The same discrepancy was analysed in detail by Sch\"ulke {\it et al.}~\cite{Schulke-Li}, where the plasmon satellite, which is present in the interacting many-body spectral function, but absent in DFT, was speculated to be the cause of density transfer from low to high momenta.
This explanation appeared promising when the $G^0W^0$ Compton profiles by Kubo~\cite{Kubo-Li-GW} seemed to resolve the low momentum discrepancies between experiment and theory. The presence of the plasmon satellite caused the redistribution of the quasiparticle occupation distribution $n^{\sigma}_{\eta\mathbf{k}}$, suppressing the $G^0W^0$ Compton profile at low momentum with respect to the corresponding LDA profile. The inclusion of many-body effects improve other aspects of the electron structure, for example, 
QSGW calculations of Li by Friedrich {\it et al.}~\cite{Friedrich-QSGW} show an improved predicted bandwidth of 3.10 eV, close to the experimental value of 3.0 eV~\cite{Vos-Li-exp-Akw}, and better than the bandwidths from $G^0W^0$ (3.22 eV) and DFT (3.46 eV) calculations. They concluded that the bandwidth overestimation of DFT is mainly due to long-range correlations which are well described by the $GW$ self-energy approximation.
Other calculations, using instead dynamical mean field theory (DMFT), which models strong local interactions and therefore suggests the opposite interpretation, also predicted improved bandwidths: DFT+eDMFT~\cite{mandal2022electronic} shows bandwidths less than 3.0 eV.

Alternatively, to model temperature effects present in room temperature experiments, Dugdale and Jarlborg pursued a theoretical approach which considered thermal effects on the Compton profile~\cite{DUGDALE1998283}, since thermal disorder was likely to be considerable at 300~K given its low atomic mass, low melting point, and low Debye temperature. 
Although their calculations predicted a broadening of the Compton profiles similar to the experimental observations, subsequent experiments~\cite{Sternemann-Li-temp-CPs,Chen-Li-temp-CPs} in fact showed an opposite trend, with the profile broadening at lower temperatures, rather than narrowing as a consequence of the reduced lattice spacing.
Quantum Monte Carlo (QMC) predictions by Filippi and Ceperley~\cite{Filippi-Li-QMC} indicate that the impact of electronic correlation in lithium only accounts only for about 30\% of the discrepancy and they point to other factors such as temperature, finite-size errors specific to the QMC approach adopted as well as errors caused by their pseudopotential method, in order to explain the difference between the experiments and their QMC calculations.  This study also concluded that thermal disorder was not the significant contributor to the discrepancy with experiment~\cite{Sternemann-Li-temp-CPs}.
These factors were revisited in detail from the theoretical perspective when Yang {\it et al.}~\cite{Yang-QMC-Li} compared their Li QMC Compton profiles with the high resolution data of polycrystalline Li from Ref.~\cite{Hiraoka-Li-Z-Compton}. 
With all the predicted results thus far, a truly compelling explanation for the discrepancies in the measurement and calculation of Compton profiles in Li is yet to be presented.

In order to determine whether $GW$ alone is sufficient to account for the discrepancy in Li Compton profiles, we compare the Li Compton profiles along high-symmetry crystallographic directions obtained in experiments by Sakurai et al.~\cite{Sakurai-Li} with those predicted by DFT, $G^0W^0$ and QSGW. These comparisons are also excellent for benchmarking Compton profiles from beyond-DFT methods such as those implemented within Questaal. 
In Fig.~\ref{fig:Li-theory-vs-experiment}, we demonstrate that the DFT Compton profiles from the Elk and Questaal codes are in overall excellent agreement with each other. We note that there are small noticeable differences between the corresponding DFT directional differences generated by Elk and Questaal, although these are not unsurprising given the different methodologies these software packages use to solve the Kohn-Sham equations.  
Looking at the insets of Fig.~\ref{fig:Li-theory-vs-experiment} and the directional differences, we see that the QSGW has a slight overall improvement with the experimental data with respect to the DFT Compton profiles.
The Elk $G^0W^0$ Compton profiles have the best agreement with the experimental data within Fig.~\ref{fig:Li-theory-vs-experiment}. However, our $G^0W^0$ Compton profiles do not reproduce the excellent experiment–theory agreement reported in the previous $GW$ calculation by Kubo~\cite{Kubo-Li-GW}. 
The discrepancy between the $GW$ results presented here to those by Kubo may be explained by the limited computational resources and the approximations used by Kubo to feasibly calculate the $GW$ EMD. 
From the Elk $G^0W^0$ Compton profiles, the explanation of the low-momentum experiment-theory discrepancies being reconciled when including the $G^0W^0$ spectral function satellite features, previously affirmed by Kubo, does not appear to be the case. 
As highlighted within the insets of Fig.~\ref{fig:Li-theory-vs-experiment}, the improved agreement between the $G^0W^0$ and experimental Compton profiles is most significant at low momentum and around the Fermi break ($\mathbf{k}_F$). This improvement originates from the many-body nature of the occupation distribution of the $G^0W^0$ quasiparticles, whereas DFT and QSGW are still use the independent particle occupation distribution. This redistribution effect of the many-body occupation distribution is also illustrated when comparing the Compton profiles of the free electron gas and interacting electron gas~\cite{cooper_compton_1997}. 
Therefore, these many-body effects are missing from the QSGW reduced density matrix and thus the Compton profiles as these are generated from the quasiparticlised Hamiltonian representing independent electron quasiparticles. 
Apart from this, the Compton profiles from the $G^0W^0$ and QSGW methods agree.

Our results highlight the sensitive nature of the Compton profiles to the influence of many-body effects in ``simple'' metals and that those captured within $GW$ methods, particularly the inclusion of satellite features, are not enough to account for the discrepancies with the Li experimental data, as has previously been claimed~\cite{Schulke-Li}.
The nature of these discrepancies still need to be resolved.

\begin{figure*}[!ht]
        \centerline{\includegraphics[width=0.93\linewidth]{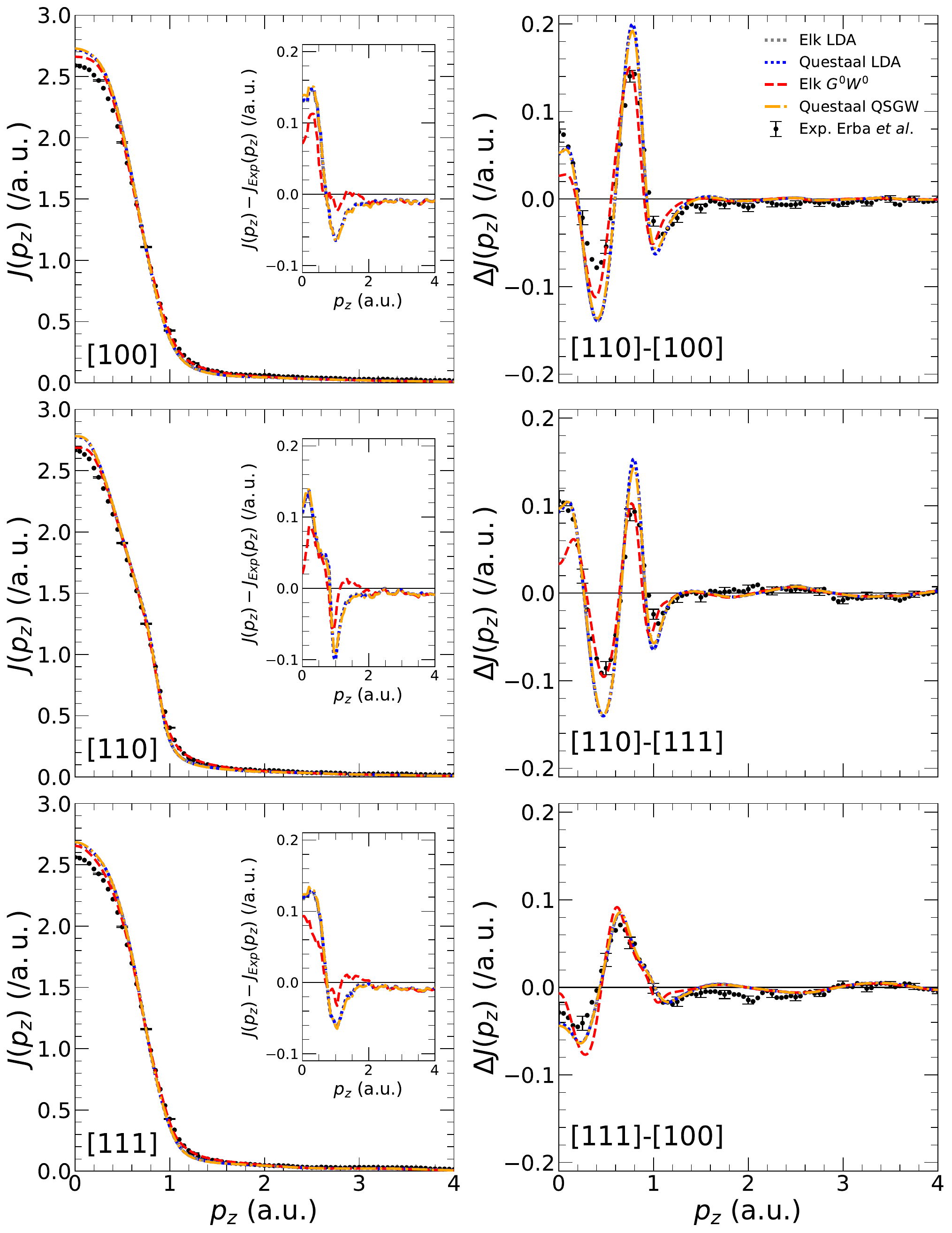}}
        \caption{
        Left column: Comparison of three high-symmetry direction ([001], [011], [111]) experimental Si Compton profiles (CPs) from Erba~\textit{et al.}~\cite{Erba_2011} (dots) with our Elk (LDA), Elk $G^0W^0$, Questaal DFT (LDA) and QSGW CPs. 
        The insets highlight the difference between the theoretical and experimental (without error bars) CPs. We note that the experimental profiles were normalised to the number of electrons the profiles represent up to 10 a.u..
        Right column: The theoretical and experimental directional differences of the high symmetry directions ([001], [011], [111]) from the left column. We show the experimental error bars for every fifth data point. 
        We convolute our calculations with a Gaussian experimental resolution function (0.11 a.u. FWHM). 
        } 
        \label{fig:Si:theory-vs-experiment}
\end{figure*}

\begin{figure*}[!ht]
        \centerline{\includegraphics[width=0.93\linewidth]{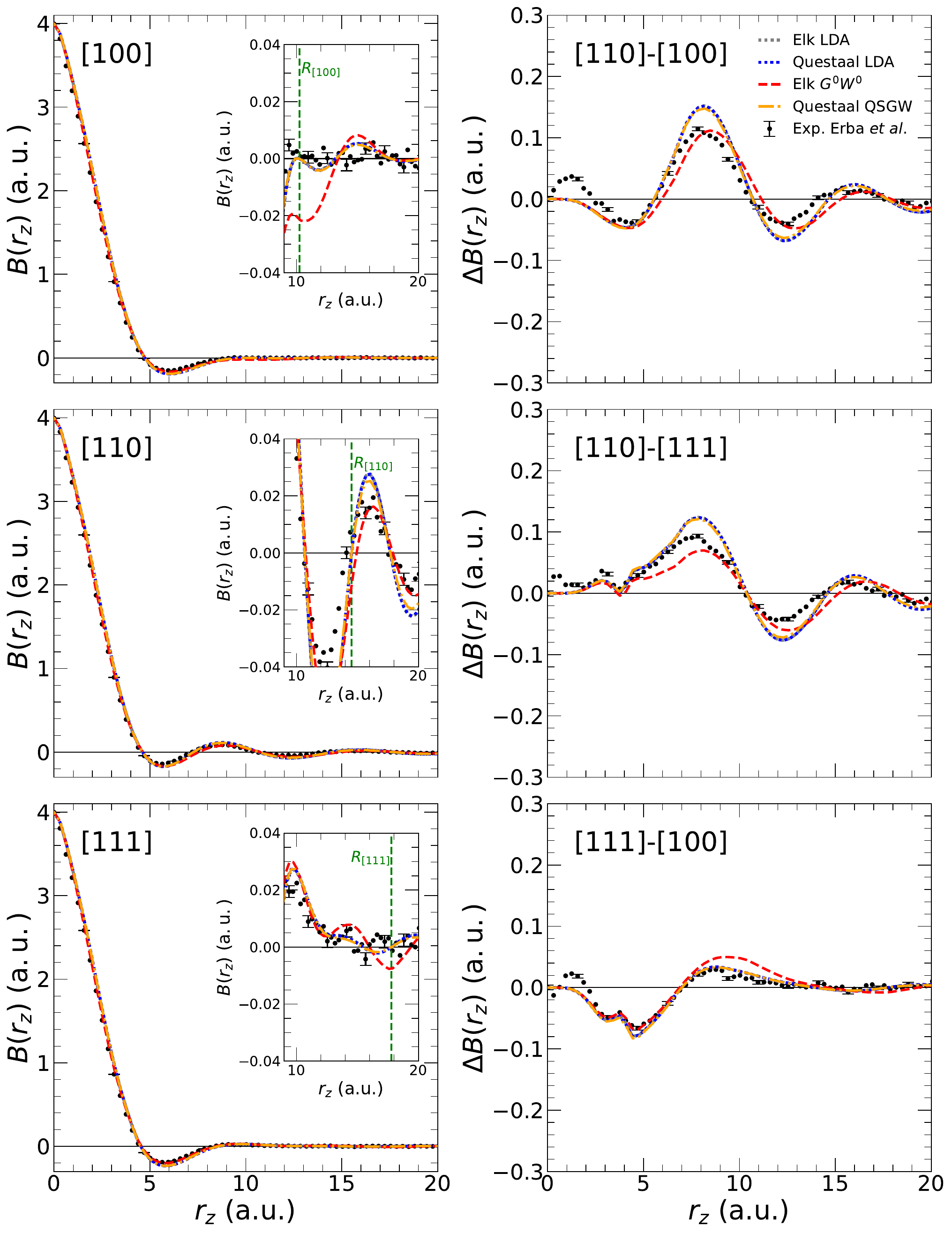}}
        \caption{Left column:  The 1D slice of the theoretical and experimental auto-correlation functions $B(r_z)$ of Si along the three high-symmetry directions ([001], [011], [111]). These were obtained by Fourier transforming the experimental and convoluted theoretical Compton profiles in Fig.~\ref{fig:Si:theory-vs-experiment}. The inset zooms into $B(r_z)$ to highlight it at the lattice vector along each high symmetry. Right column: the auto-correlation function anisotropies between the different high symmetry directions. The presented $B(r_z)$ errors were calculated from the standard deviation of the Fourier transform of 10$^3$ experimental Compton profiles each with added random noise based on a normal distribution with the experimental Compton profile error. 
        }
        \label{fig:Si:Br:theory-vs-experiment}  
\end{figure*}

\subsection{Si}

Semiconducting Si has been intensively studied by Compton scattering~\cite{Weiss_1972,shulke_1974,Seth_1977,Pattison_1978,Pattison_1981,hansen_1987,Nara_1979,Nara_1984,Sakai_1989,Kubo_1997,Delaney_1998,Tse_2005,Erba_2011,Pisani_2011,Okada_2012,Matsuda_2013,Klevak_2016,Aguiar_2026}. The focus of most of the early studies was the influence the Si bonding had on the Compton profile, and often focused on the anisotropy (differences between profiles measured for different crystallographic directions) ~\cite{hansen_1987,Pattison_1981,Sakai_1989,Nara_1979,Nara_1984}. Insight into the bonding came from both the auto-correlation function~\cite{hansen_1987,Pattison_1981} or the Compton profiles themselves~\cite{Sakai_1989,Nara_1979,Nara_1984}. 
More recently, Compton scattering has been used to extract the electronic structure of Si under high pressure~\cite{Tse_2005} and even liquid Si~\cite{Okada_2012,Matsuda_2013,Klevak_2016}. Furthermore, there has been interest in resolving the (low momentum) discrepancies between the theoretical and experimental Compton profiles~\cite{Erba_2011,Pisani_2011,Aguiar_2026}. Indeed, Kubo \textit{et al.}~\cite{Kubo_1997} looked at including the self-interaction correction (SIC) in their calculations, but this yielded little improvement. The more recent, closely related studies~\cite{Erba_2011,Pisani_2011} have used Hartree-Fock based methods alongside the DFT-based Compton profiles and found that the post-Hartree-Fock method gave improved agreement, especially when comparing to the (1D slice) $B(\textbf{r})$ at $B(\textbf{R}) = 0$. The conclusions of these studies are that the discrepancies may be attributed to the neglect of the nuclear motion and/or the theoretical treatment, where the post-Hartree-Fock results indicate that the instantaneous electron correlation effects in the Compton profiles cannot be reproduced from the reduced density matrix obtained from a single-determinantal wavefunction. 

$G^0W^0$ and QSGW are renowned for their success in drastically improving the band gap across a plethora of materials, including Si, compared with LDA and GGA~\cite{van_schilfgaarde_adequacy_2006,Faleev_2006,kotani_quasiparticle_2007,Grumet_2018,Salas-Illanes_2022}. With respect to Si Compton profiles, QSGW is uniquely positioned to test whether the improvement from the $GW$ electron correlations within the single-particle framework can resolve the discrepancies, or whether a many-body treatment is more likely needed for Si. 
The presence of the low momentum discrepancy between the DFT and experimental Compton profiles and directional differences within Fig.~\ref{fig:Si:theory-vs-experiment} are in agreement with that previously seen~\cite{Kubo_1997,Delaney_1998,Erba_2011,Pisani_2011}. We also highlight the excellent agreement between the Questaal and Elk DFT Compton profiles and directional differences. The QSGW Compton profiles and directional differences offer little to reduce the discrepancies. 
Despite this, these theories do predict the directional differences reasonably well, especially along the [111]-[100] direction. 
Clearly, the improvements predicted by QSGW for the Si electronic structure do not affect the Compton scattering experiment–theory discrepancies. However, this is a very different story for $G^0W^0$ as its Compton profiles and directional differences are drastically different compared to the DFT and QSGW ones.
Indeed, it appears that the $G^0W^0$ results have better agreement at several momentum regions with respect to the other theoretical methods, although its directional differences discrepancies tend to be worse at lower momentum. The fact that both $G^0W^0$ and QSGW have similar improvements to the Si electron structure, such as the band gap, indicates that the differences between the Compton profiles of the two methods are unlikely related to the single-body exchange-correlation description (the band gap will unlikely change the EMD for large gaps).

For completeness, we present the 1D slice of the auto-correlation function $B(r_z)$ and its directional anisotropies within Fig.~\ref{fig:Si:Br:theory-vs-experiment}. Again, we see similar behaviour between the DFT and QSGW $B(r_z)$, with drastic differences within the $G^0W^0$. It is clear that the $G^0W^0$ directional anisotropies has significantly worse agreement with the experimental data with respect to the other theoretical predictions. The origin of these disagreements can be seen from closer inspection of the $B(r_z)$, such as those shown within the insets of Fig.~\ref{fig:Si:Br:theory-vs-experiment}. It is important to reiterate that, within the single-particle approximation, the $B(\mathbf{r})$ function in an insulator such as Si is expected to vanish at real-space lattice translation vectors ($B(\mathbf{R}) = 0$) as a consequence of translational symmetry and the orthogonality of the occupied Bloch orbitals~\cite{cooper:85}. Here, both DFT and QSGW obey this $B(\mathbf{R}) = 0$ condition due to their independent-particle nature and the absence of a Fermi surface. This condition is also reasonably well reproduced in the experimental data, although the point at which the experimental $B(\mathbf{r})$ crosses zero deviates by a small but notable amount from the lattice translation vectors $\mathbf{R}$. This is likely due to a small but non-negligible breakdown of the single-particle approximation. Therefore, the $B(\mathbf{R}) = 0$ condition is expected to be relatively robust for Si. However, $G^0W^0$ clearly and significantly deviates from this condition, primarily because its quasiparticles are no longer constrained by the single-particle approximation. It should be noted that the post–Hartree–Fock results in Refs.~\cite{Erba_2011,Pisani_2011} do not significantly break this condition and are in fact in closer agreement with the experimental $B(\mathbf{r})$ around $\mathbf{R}$. This suggests that many-body $GW$ electron-correlation effects, rather than exchange interactions, are responsible for the breakdown of this condition. Therefore, it is clear that there are still missing electron-correlation contributions within the $G^0W^0$ framework that are required to correct $B(\mathbf{r})$ and thus reduce this significant deviation from the $B(\mathbf{R}) = 0$ condition. 

From Figs.~\ref{fig:Si:theory-vs-experiment} and \ref{fig:Si:Br:theory-vs-experiment}, the changes between the $G^0W^0$ and QSGW Compton profiles and auto-correlation functions are most likely due to the approach in computing the reduced density matrix. The QSGW method uses its reduced density matrix from independent Kohn-sham orbitals to calculate the EMD whereas $G^0W^0$ uses its many-body reduced density matrix.  This ties in well with the previous post-Hartree-Fock results~\cite{Erba_2011,Pisani_2011} where it was proposed that the origin of the discrepancies appear to be beyond a one-particle treatment of the EMD. Our results confirm that this is indeed a significant factor.

\subsection{Cr}

\begin{figure*}[!ht]
        \centerline{\includegraphics[width=0.88\linewidth]{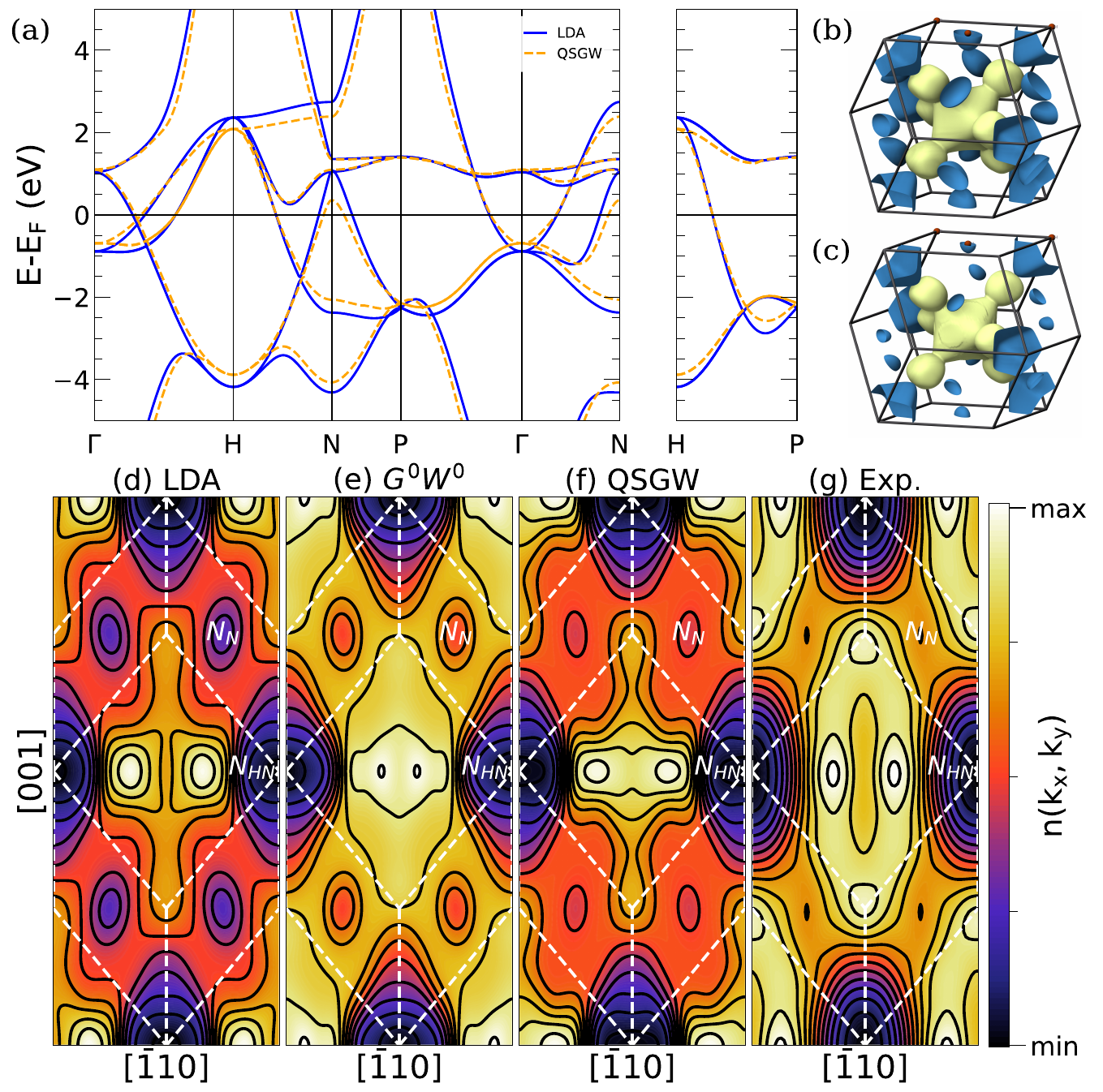}}
        \caption{(a) The single-particle non-magnetic Cr LDA and Questaal QSGW band structures along the same path as presented in Ref.~\cite{harris-lee_sensitivity_2021}. The corresponding Fermi surface of Cr as described by the (b) LDA and (c) QSGW. 
        The 2D (once projected) occupation densities n($\rm k_x, k_y$) along the [110] high symmetry direction from (c) LDA, (d) Elk $G^0W^0$, (e) Questaal QSGW and (e) the reconstruction from Dugdale~\textit{et al.}~\cite{tanaka_study_2000,DUGDALE2000361}. These theoretical n($\rm k_x, k_y$) are non-magnetic. We note that the presented LDA plots are from Questaal but Elk gives consistent plots for its LDA results.
        We convolute our n($\rm k_x, k_y$) calculations with a 2D Gaussian (0.18 a.u. FWHM) to simulate the experimental resolution.
        Some projected symmetry points in the projected Brillouin zone are labelled. 
        }
        \label{fig:Cr:LCWs}
\end{figure*}

\begin{figure*}[!ht]
        \centerline{\includegraphics[width=0.95\linewidth]{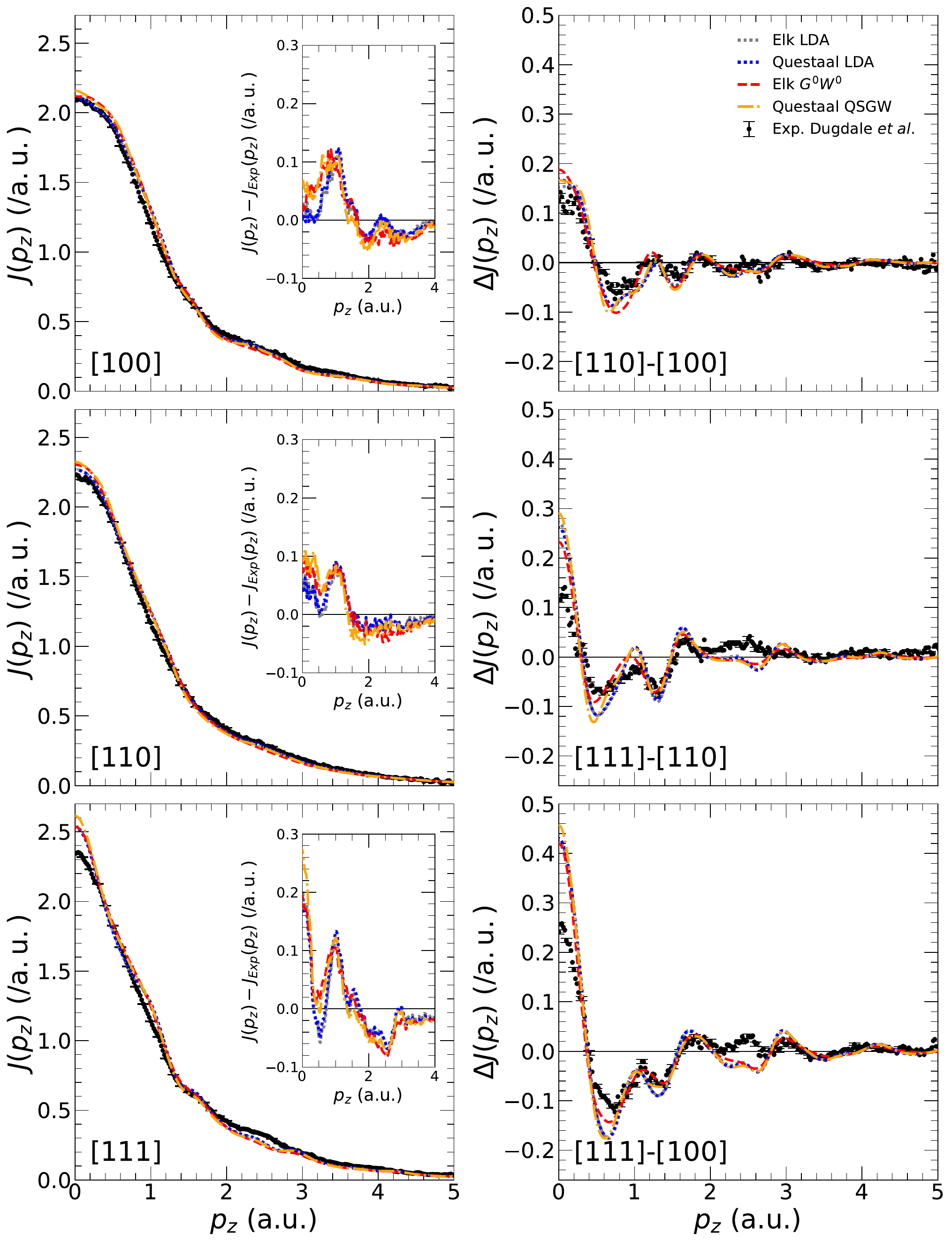}}
        \caption{
        Left column: Comparison of three high-symmetry direction ([001], [011], [111]) experimental Cr Compton profiles (CPs) from Dugdale~\textit{et al.}~\cite{tanaka_study_2000,DUGDALE2000361} (dots with estimated error bars) with our non-magnetic Questaal LDA, Elk LDA, Elk $G^0W^0$ and QSGW CPs. 
        The insets highlight the difference between the theoretical and experimental (without error bars) CPs. Right column: The theoretical and experimental directional differences of the high symmetry directions ([001], [011], [111]) from the left column. 
        The experimental error has been plotted for every fifth data point.
        We convolute our calculations with an experimental resolution function (0.18 a.u. FWHM).
        } 
        \label{fig:Cr:theory-vs-experiment}
\end{figure*} 

Cr has long stood out among the early 3d and 4d bcc transition metals (which include V, Nb, Cr, and Mo) because it exhibits an incommensurate spin-density wave below its N\'eel temperature of 311~K~\cite{RevModPhys.60.209} while the other elements are free of magnetic order.
Although the experimentally measured Fermi surfaces of V, Nb, and Mo are rather well described by DFT calculations, the predictions for paramagnetic Cr have been markedly inferior, particularly in describing the $N$-hole ellipsoids (see~\cite{harris-lee_sensitivity_2021} and references therein); these originate in a band with dominant $p$ character and have always appeared smaller than the DFT predictions in both momentum distributions obtained from Compton scattering and positron annihilation experiments of paramagnetic Cr~\cite{Fretwell_1995,matsumoto_86,dugdale_fermiology_1998,tanaka_study_2000,DUGDALE2000361,PhysRevB.69.174406}.
The calculated Cr Fermi surface depends sensitively on the accuracy of the treatment of electronic correlation and magnetism, and conventional DFT calculations are not satisfactory: experimental insight from de Haas-van Alphen (dHvA) effect \cite{graebner_haas-van_1968,laurent_band_1981} (probing the Cr anti-ferromagnetic phase), two-dimensional angular correlation of positron annihilation radiation (2D-ACAR) \cite{dugdale_fermiology_1998}, and Compton scattering \cite{tanaka_study_2000,DUGDALE2000361} all consistently indicate a Fermi surface that is quantitatively different from the existing theoretical predictions, including DFT calculations with the local density approximation (LDA) \cite{laurent_band_1981,kubler_spin-density_1980}.
Indeed Haris-Lee~\textit{et al.}~\cite{harris-lee_sensitivity_2021} showed that the exchange-correlation treatment within QSGW and metaGGA is important in producing better agreement with the those aforementioned experimental probes. However, the QSGW Compton profiles and 2D (once projected) occupation distributions were not presented as those non-magnetic calculations were not possible at the time; we present them here.

Fig.~\ref{fig:Cr:LCWs}~(a) shows the non-magnetic LDA and QSGW band structure which has good agreement with that plotted along the same path in Ref.~\cite{harris-lee_sensitivity_2021}. It is clear that the most significant impact of the QSGW is on the band near the Fermi level around the $N$ point. The corresponding LDA and QSGW Fermi surfaces are shown in Fig.~\ref{fig:Cr:LCWs} (b) and (c), respectively, which clearly show the shrinking of the $N$-hole ellipsoids. These shrunken ellipsoids are also visible within the (projected) 2D occupation distribution, the experimental 2D occupation distribution having been reconstructed from the Compton scattering measurements~\cite{tanaka_study_2000,DUGDALE2000361}. Figs.~\ref{fig:Cr:LCWs} (d)-(f) show the non-magnetic theoretical LDA, $G^0W^0$ from the LDA starting point, and QSGW 2D occupation distributions, respectively. Lastly, Fig.~\ref{fig:Cr:LCWs} (g) shows the paramagnetic experimental 2D occupation distribution. The QSGW nicely captures the reduction of the $N$-hole ellipsoids (with respect to the LDA), see the Fermi surfaces in Figs.~\ref{fig:Cr:LCWs} (b) and (c), which better agrees with the measured 2D occupation distribution when comparing these ellipsoids within Figs.~\ref{fig:Cr:LCWs} (d)-(g). As previously stated, this improvement of the ellipsoids in QSGW originates from QSGW better treating the $p$ character dominant band around $N$ point.
Interestingly, the central region of the $G^0W^0$ 2D occupation distribution, which is associated to the projection of the so-called ``jack'' Fermi surface in Figs.~\ref{fig:Cr:LCWs} (b) and (c), has a superior agreement with the experimental data compared to the other theories. This is likely attributable to the many-body reduced matrix being used instead of the single-particle one used by the other theories. This further highlights the importance of the many-body reduced density matrix for capturing the smearing of the occupation distribution observed experimentally. However, in Fig.~\ref{fig:Cr:LCWs} (e) it appears that the $N$-hole pocket regions in the $G^0W^0$ 2D occupation distribution are still a bit more pronounced, meaning worse agreement than that predicted by QSGW. We note that the $N$-hole pocket is particularly sensitive to calculation parameters for one-shot $GW$ - for example, the first QSGW iteration band structure presented by Haris-Lee~\textit{et al.}~\cite{harris-lee_sensitivity_2021} showed no $N$ hole pocket Fermi sheet, which is different from that shown in the Elk $G^0W^0$ 2D occupation distribution within Fig.~\ref{fig:Cr:LCWs} (e). 

The QSGW Compton profiles and directional differences appear to have slight worse agreement with the experimental data within Fig.~\ref{fig:Cr:theory-vs-experiment} compared to the LDA counterparts. We note the LDA Compton profiles and directional differences from Elk and Questaal are consistent with each other. The insets show the worsening of the discrepancies between experiment and theory across the different theoretical frameworks. Despite this, the theories do a good job of capturing the anisotropy along the [110]-[100] directions, but it is clear that the discrepancies are the worst along the [111] direction. Interestingly, the $G^0W^0$ Compton profiles and directional differences have a better overall agreement with the corresponding experimental data with respect to the QSGW results. Again, as this is primarily associated with the use of the many-body reduced density matrix, it further strengthens our argument for its importance. 

\begin{figure*}[!ht]
        \centerline{\includegraphics[width=0.95\linewidth]{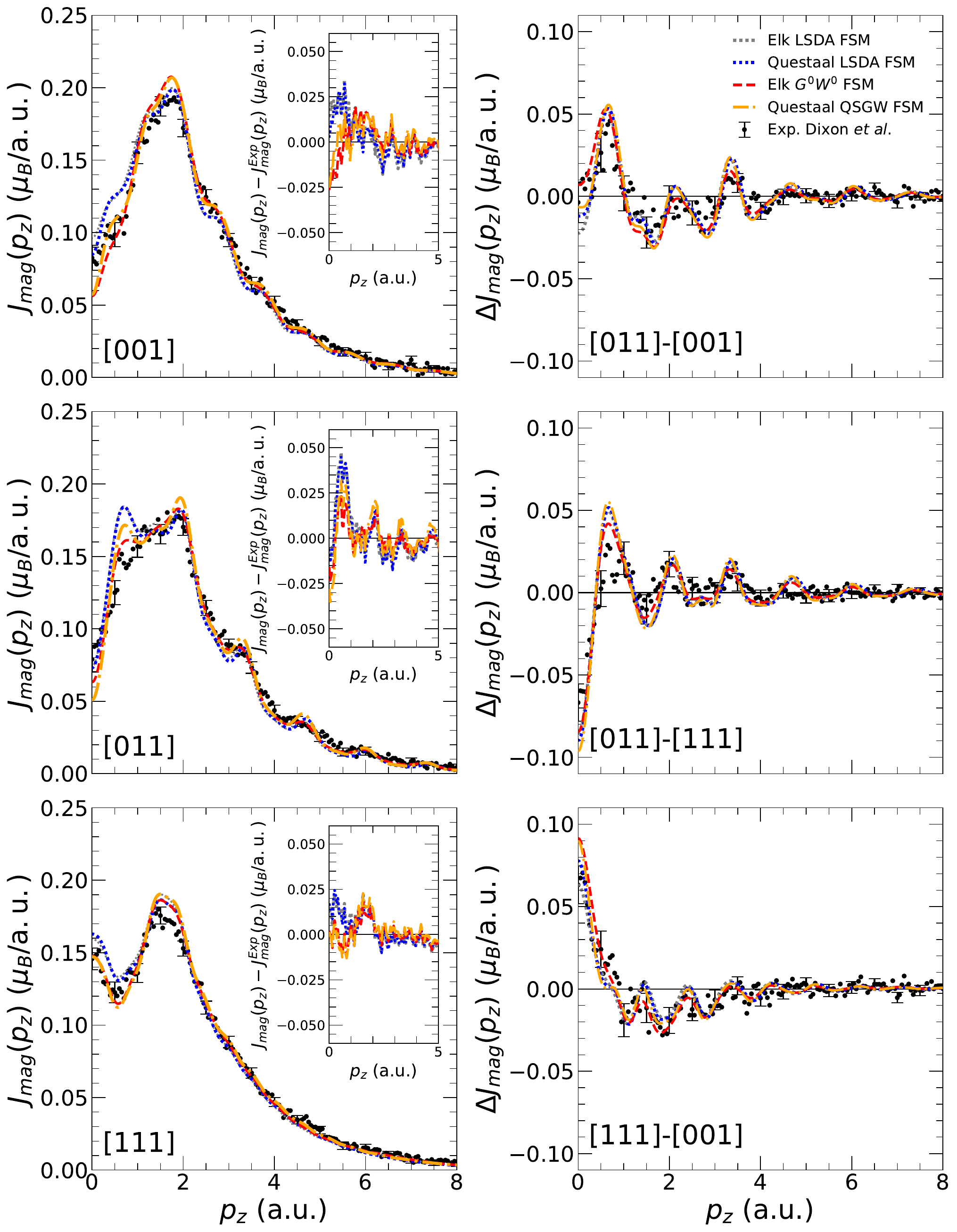}}
        \caption{
        Left Column: Comparison of three high-symmetry direction ([001], [011], [111]) experimental Ni magnetic Compton profiles (MCPs) from Dixon~\textit{et al.} ~\cite{di.du.98} (dots with error bars) with our fixed-spin moment (FSM) DFT from Elk and Questaal, Elk $G^0W^0$ and Questaal QSGW MCPs. 
        The insets show the differences between the theory and the experimental data (without error bars) along each high symmetry direction. Right column: for completeness, we included the magnetic directional difference of the high symmetry directions from the left column. 
        We show the experimental error bars for every tenth data point. 
        We find that beyond 8~a.u. the experimental and theoretical MCPs are not distinguishable.
        We convolute our calculations with an experimental resolution function (0.43 a.u. FWHM). 
        } 
        \label{fig:Ni:FSM-theory-vs-experiment}
\end{figure*}

\subsection{Ni}

The model material for magnetic Compton scattering studies is ferromagnetic nickel with a large and growing literature of experimental and theoretical studies~\cite{sakai_application_1991,timms_spin_1990,billington_2020}.
We note that the Lam-Platzman correction is isotropic in spin so it does not contribute within magnetic Compton profiles (MCPs). Therefore experiment-theory comparisons are sensitive to the size and localisation of the magnetic moment.
The high quality MCP experimental data available for Ni make it an excellent material with which to benchmark theoretical techniques~\cite{James_2025}.
The low momentum (up to $\sim 3$~a.u.) region of the MCP is particularly poorly described by DFT~\cite{di.du.98} and Kubo applied $G^0W^0$ to the Ni already in 2004~\cite{kubo_electron_2004} only to find that the improved theory actually gave a worse description. 
The $G^0W^0$ MCP calculations by James et al.~\cite{James_2025} extended those of Kubo, in particular by showing that $G^0W^0$ Ni MCPs constrained to the experimental spin moment show excellent agreement with experiment. 
Indeed, the MCPs obtained by theoretical methods including DFT, $G^0W^0$, and DFT+DMFT showed that the addition of both a fully non-local treatment (at the $GW$ level) of electron correlations and a treatment of fluctuations of local spin present in DMFT for uniformly reducing the area under the MCP  (which is, of course, the spin moment) are needed to correct for the discrepancies present between LSDA/GGA MCPs and experimental data.  
In fact, this agreed with the conclusions from Ref.~\cite{sponza_2017} where their QSGW+DMFT and fixed-moment QSGW results yielded good agreement of both the spin splitting and the $d$-band width with ARPES data.

James et al.~\cite{James_2025} noted that an improved description of spin polarization in the $G^0W^0$ calculations is key in producing the remarkable agreement in the MCP shape with the experimental data in the low momentum region.
This improvement cannot be achieved within DFT or the purely local modifications in DFT+DMFT.
The smearing of the occupation distribution in $G^0W^0$ calculations is an important factor in getting such an excellent agreement of the shape between the fixed-spin moment (FSM) $G^0W^0$ and experimental MCPs: in particular, the $G^0W^0$ occupation distribution suppresses the so-called ``Umklapp'' peaks in the MCPs (which are simply a consequence of the Bloch nature of the wavefunctions in the extended zone scheme).
The inclusion of the 6 eV satellite, which is well described in DMFT, has no significant influence on the MCPs, but the $G^0W^0$ diagonal approximation (limiting the effect of $GW$ to eigenvalue shifts) predict MCPs with significantly worse agreement with the experimental Compton data, especially in the low momentum region. 
However, these conclusions are in part based on the $G^0W^0$ method which may retain influence from the initial LSDA calculation used, as suggested by James et al.~\cite{James_2025}.
MCPs calculated from FSM QSGW should highlight whether these conclusions are robust and independent to the $GW$ method used.

The FSM DFT MCPs from Elk and Questaal in Fig.~\ref{fig:Ni:FSM-theory-vs-experiment} are in excellent agreement. 
With this confirmed, we can compare the MCPs from the $GW$ methods. Overall, Elk's $G^0W^0$ and Questaal's QSGW MCPs shows good agreement with each other, especially in resolving the low momentum discrepancies with the experimental data.
Clearly this helps to verify the conclusions of James et al.~\cite{James_2025} and indicates that the improved agreement from the $GW$ MCPs are independent of starting from LSDA or the self-consistent converged QSGW potential in this case.
Interestingly, the largest discrepancies between the the $GW$ methods are at the Umklapp peaks in the MCPs, which is clearly seen in the [011] direction.
Along the [111] direction where there are no clear Umklapp peaks, the agreement between the $GW$ methods is excellent.
As shown by James et al.~\cite{James_2025}, these Umklapp peaks are suppressed by the influence of the many-body occupation distribution. The absence of such suppression in the QSGW MCPs further supports this interpretation, since the MCPs are generated within a single-particle QSGW framework. 
This interacting occupation distribution and its effect of suppressing these peaks are directly linked to the many-body nature of the reduced density matrix used within $G^0W^0$ and therefore highlights its importance.
Clearly, the QSGW shows that the improvement in the low momentum region can be captured within the quasiparticlised potential. For completeness, we included the magnetic directional difference of the high symmetry directions. We see that the there is good agreement between all theoretical and the experimental differences but the errors are too large to get any meaningful interpretation of any improvements within the theory.
From Fig.~\ref{fig:Ni:FSM-theory-vs-experiment}, these MCPs help to reinforce the conclusions of James et al.~\cite{James_2025} by reproducing the improved low momentum agreement from an alternative $GW$ method although the discrepancies between the $GW$ methods are attributed to the single-particle nature of the QSGW reduced density matrix. 

\section{Conclusion}
We have presented the Compton profiles of several elements from the Questaal package including those from the QSGW method. We see that the QSGW Compton profiles are consistent with the many-body $G^0W^0$ results for Li and Si, where the differences in the profiles from the $GW$ methods are attributed to the single-particle nature of the QSGW reduced density matrix used for calculating the Compton profiles. For Li, the our $G^0W^0$ and QSGW results confirm that the $GW$ method is unable to resolve the experiment-theory discrepancies, contrary to that reported by Kubo~\cite{Kubo-Li-GW}, which means these discrepancies require further investigation. The QSGW Compton profiles of semi-conducting Si show no significant improvements compared to the experimental profiles with respect to the corresponding DFT profiles, despite the improvements QSGW gives for other aspects of the Si electronic structure. This is not unexpected as there is evidence that the discrepancies may relate to the single-particle nature of the reduced density matrix~\cite{Pisani_2011} which is used by QSGW. This is further compounded by the drastic changes in the EMD and auto-correlation function from $G^0W^0$ which uses its many-body reduced density matrix to compute the EMD. 

Furthermore, we present the QSGW Compton profiles and 2D projected occupation distribution of Cr, partly motivated by QSGW doing a very good job of improving the Fermiology, especially around the $N$-hole pockets~\cite{harris-lee_sensitivity_2021}. Indeed, we see improved signatures of the $N$-hole pockets between the QSGW and experimental data. However, our many-body $G^0W^0$ again highlights the need for the many-body reduced density matrix for improved agreement with the experiment. Finally, the QSGW magnetic Compton profiles Ni, on the other hand, yield excellent agreement with the corresponding $G^0W^0$ and experimental profiles. Indeed, the Umklapp peaks are more prominent within the QSGW magnetic Compton profiles with respect to the $G^0W^0$ and experimental magnetic Compton profiles, again due to the single-particle nature of the QSGW reduced density matrix, which was expected~\cite{James_2025}. 

These comparisons of state-of-the-art theoretical methods with Compton scattering data prove insightful to evaluate the quality of the theory and gain further key insights. 
Our results demonstrate that accurate quasiparticle electronic structures do not necessarily imply accurate electron momentum densities. While QSGW significantly improves many aspects of electronic structure, the EMD remains highly sensitive to the many-body structure of the reduced density matrix and the redistribution of occupation weight arising from electron correlations. Indeed, our results provide motivation for incorporating the many-body reduced density matrix into the QSGW framework. Compton scattering therefore provides a uniquely useful probe of many-body electronic structure methods beyond conventional band-structure validation.

\section{Acknowledgements}
JJ acknowledges support from CCP9 ``Computational Electronic Structure of Condensed Matter'', part of the Computational Science Centre for Research Communities (CoSeC). Calculations were performed using the computational facilities of the Advanced Computing Research Centre, University of Bristol (\href{http://bris.ac.uk/acrc/}{http://bris.ac.uk/acrc/}) along with computing resources provided by STFC Scientific Computing Department’s SCARF cluster (\url{https://www.scarf.rl.ac.uk/}).


\begin{thebibliography}{81}%
\makeatletter
\providecommand \@ifxundefined [1]{%
 \@ifx{#1\undefined}
}%
\providecommand \@ifnum [1]{%
 \ifnum #1\expandafter \@firstoftwo
 \else \expandafter \@secondoftwo
 \fi
}%
\providecommand \@ifx [1]{%
 \ifx #1\expandafter \@firstoftwo
 \else \expandafter \@secondoftwo
 \fi
}%
\providecommand \natexlab [1]{#1}%
\providecommand \enquote  [1]{``#1''}%
\providecommand \bibnamefont  [1]{#1}%
\providecommand \bibfnamefont [1]{#1}%
\providecommand \citenamefont [1]{#1}%
\providecommand \href@noop [0]{\@secondoftwo}%
\providecommand \href [0]{\begingroup \@sanitize@url \@href}%
\providecommand \@href[1]{\@@startlink{#1}\@@href}%
\providecommand \@@href[1]{\endgroup#1\@@endlink}%
\providecommand \@sanitize@url [0]{\catcode `\\12\catcode `\$12\catcode `\&12\catcode `\#12\catcode `\^12\catcode `\_12\catcode `\%12\relax}%
\providecommand \@@startlink[1]{}%
\providecommand \@@endlink[0]{}%
\providecommand \url  [0]{\begingroup\@sanitize@url \@url }%
\providecommand \@url [1]{\endgroup\@href {#1}{\urlprefix }}%
\providecommand \urlprefix  [0]{URL }%
\providecommand \Eprint [0]{\href }%
\providecommand \doibase [0]{https://doi.org/}%
\providecommand \selectlanguage [0]{\@gobble}%
\providecommand \bibinfo  [0]{\@secondoftwo}%
\providecommand \bibfield  [0]{\@secondoftwo}%
\providecommand \translation [1]{[#1]}%
\providecommand \BibitemOpen [0]{}%
\providecommand \bibitemStop [0]{}%
\providecommand \bibitemNoStop [0]{.\EOS\space}%
\providecommand \EOS [0]{\spacefactor3000\relax}%
\providecommand \BibitemShut  [1]{\csname bibitem#1\endcsname}%
\let\auto@bib@innerbib\@empty
%</preamble>
\bibitem [{\citenamefont {Cooper}(2004)}]{cooper_x-ray_2004}%
  \BibitemOpen
  \bibfield  {author} {\bibinfo {author} {\bibfnamefont {M.~J.}\ \bibnamefont {Cooper}},\ }\href@noop {} {\emph {\bibinfo {title} {X-ray {Compton} scattering}}}\ (\bibinfo  {publisher} {OUP Oxford},\ \bibinfo {year} {2004})\BibitemShut {NoStop}%
\bibitem [{\citenamefont {Dugdale}(2014)}]{dugdale:14}%
  \BibitemOpen
  \bibfield  {author} {\bibinfo {author} {\bibfnamefont {S.~B.}\ \bibnamefont {Dugdale}},\ }\bibfield  {title} {\bibinfo {title} {Probing the {F}ermi surface by positron annihilation and {C}ompton scattering},\ }\href {https://doi.org/10.1063/1.4869588} {\bibfield  {journal} {\bibinfo  {journal} {Low Temperature Physics}\ }\textbf {\bibinfo {volume} {40}},\ \bibinfo {pages} {328} (\bibinfo {year} {2014})}\BibitemShut {NoStop}%
\bibitem [{\citenamefont {Cooper}\ and\ \citenamefont {Duffy}(2000)}]{cooper_spin_2000}%
  \BibitemOpen
  \bibfield  {author} {\bibinfo {author} {\bibfnamefont {M.~J.}\ \bibnamefont {Cooper}}\ and\ \bibinfo {author} {\bibfnamefont {J.~A.}\ \bibnamefont {Duffy}},\ }\bibfield  {title} {\bibinfo {title} {Spin densities studied in momentum space},\ }\href {https://doi.org/10.1016/S0022-3697(99)00314-5} {\bibfield  {journal} {\bibinfo  {journal} {J. Phys. Chem. Solids}\ }\textbf {\bibinfo {volume} {61}},\ \bibinfo {pages} {345} (\bibinfo {year} {2000})}\BibitemShut {NoStop}%
\bibitem [{\citenamefont {Barbiellini}(2013)}]{barbiellini:13}%
  \BibitemOpen
  \bibfield  {author} {\bibinfo {author} {\bibfnamefont {B.}~\bibnamefont {Barbiellini}},\ }\bibfield  {title} {\bibinfo {title} {High-temperature cuprate superconductors studied by x-ray {C}ompton scattering and positron annihilation spectroscopies},\ }\href {https://doi.org/10.1088/1742-6596/443/1/012009} {\bibfield  {journal} {\bibinfo  {journal} {Journal of Physics: Conference Series}\ }\textbf {\bibinfo {volume} {443}},\ \bibinfo {pages} {012009} (\bibinfo {year} {2013})}\BibitemShut {NoStop}%
\bibitem [{\citenamefont {Ruotsalainen}\ \emph {et~al.}(2018)\citenamefont {Ruotsalainen}, \citenamefont {Inkinen}, \citenamefont {Pylkk{\"a}nen}, \citenamefont {Buslaps}, \citenamefont {Hakala}, \citenamefont {H{\"a}m{\"a}l{\"a}inen},\ and\ \citenamefont {Huotari}}]{ruotsalainen2018isotropic}%
  \BibitemOpen
  \bibfield  {author} {\bibinfo {author} {\bibfnamefont {K.~O.}\ \bibnamefont {Ruotsalainen}}, \bibinfo {author} {\bibfnamefont {J.}~\bibnamefont {Inkinen}}, \bibinfo {author} {\bibfnamefont {T.}~\bibnamefont {Pylkk{\"a}nen}}, \bibinfo {author} {\bibfnamefont {T.}~\bibnamefont {Buslaps}}, \bibinfo {author} {\bibfnamefont {M.}~\bibnamefont {Hakala}}, \bibinfo {author} {\bibfnamefont {K.}~\bibnamefont {H{\"a}m{\"a}l{\"a}inen}},\ and\ \bibinfo {author} {\bibfnamefont {S.}~\bibnamefont {Huotari}},\ }\bibfield  {title} {\bibinfo {title} {The isotropic {C}ompton profile difference across the phase transition of {VO}{$_2$}},\ }\href {https://doi.org/https://doi.org/10.1140/epjb/e2018-90121-x} {\bibfield  {journal} {\bibinfo  {journal} {The European Physical Journal B}\ }\textbf {\bibinfo {volume} {91}},\ \bibinfo {pages} {225} (\bibinfo {year} {2018})}\BibitemShut {NoStop}%
\bibitem [{\citenamefont {Billington}\ \emph {et~al.}(2015)\citenamefont {Billington}, \citenamefont {Ernsting}, \citenamefont {Millichamp}, \citenamefont {Lester}, \citenamefont {Dugdale}, \citenamefont {Kersh}, \citenamefont {Duffy}, \citenamefont {Giblin}, \citenamefont {Taylor}, \citenamefont {Manuel} \emph {et~al.}}]{billington2015magnetic}%
  \BibitemOpen
  \bibfield  {author} {\bibinfo {author} {\bibfnamefont {D.}~\bibnamefont {Billington}}, \bibinfo {author} {\bibfnamefont {D.}~\bibnamefont {Ernsting}}, \bibinfo {author} {\bibfnamefont {T.~E.}\ \bibnamefont {Millichamp}}, \bibinfo {author} {\bibfnamefont {C.}~\bibnamefont {Lester}}, \bibinfo {author} {\bibfnamefont {S.~B.}\ \bibnamefont {Dugdale}}, \bibinfo {author} {\bibfnamefont {D.}~\bibnamefont {Kersh}}, \bibinfo {author} {\bibfnamefont {J.~A.}\ \bibnamefont {Duffy}}, \bibinfo {author} {\bibfnamefont {S.~R.}\ \bibnamefont {Giblin}}, \bibinfo {author} {\bibfnamefont {J.~W.}\ \bibnamefont {Taylor}}, \bibinfo {author} {\bibfnamefont {P.}~\bibnamefont {Manuel}}, \emph {et~al.},\ }\bibfield  {title} {\bibinfo {title} {Magnetic frustration, short-range correlations and the role of the paramagnetic {F}ermi surface of {PdCrO}{$_2$}},\ }\href {https://doi.org/https://doi.org/10.1038/srep12428} {\bibfield  {journal} {\bibinfo  {journal} {Scientific reports}\ }\textbf {\bibinfo {volume} {5}},\ \bibinfo {pages}
  {12428} (\bibinfo {year} {2015})}\BibitemShut {NoStop}%
\bibitem [{\citenamefont {James}\ \emph {et~al.}(2023)\citenamefont {James}, \citenamefont {Billington},\ and\ \citenamefont {Dugdale}}]{James_2023}%
  \BibitemOpen
  \bibfield  {author} {\bibinfo {author} {\bibfnamefont {A.~D.~N.}\ \bibnamefont {James}}, \bibinfo {author} {\bibfnamefont {D.}~\bibnamefont {Billington}},\ and\ \bibinfo {author} {\bibfnamefont {S.~B.}\ \bibnamefont {Dugdale}},\ }\bibfield  {title} {\bibinfo {title} {{Impact of electron correlations on the $\textbf{k}$-resolved electronic structure of PdCrO$_2$ revealed by Compton scattering}},\ }\href {https://doi.org/10.1088/2516-1075/acd28d} {\bibfield  {journal} {\bibinfo  {journal} {Electronic Structure}\ }\textbf {\bibinfo {volume} {5}},\ \bibinfo {pages} {025002} (\bibinfo {year} {2023})}\BibitemShut {NoStop}%
\bibitem [{\citenamefont {Sakurai}\ \emph {et~al.}(1995{\natexlab{a}})\citenamefont {Sakurai}, \citenamefont {Tanaka}, \citenamefont {Bansil}, \citenamefont {Kaprzyk}, \citenamefont {Stewart}, \citenamefont {Nagashima}, \citenamefont {Hyodo}, \citenamefont {Nanao}, \citenamefont {Kawata},\ and\ \citenamefont {Shiotani}}]{sakurai:95}%
  \BibitemOpen
  \bibfield  {author} {\bibinfo {author} {\bibfnamefont {Y.}~\bibnamefont {Sakurai}}, \bibinfo {author} {\bibfnamefont {Y.}~\bibnamefont {Tanaka}}, \bibinfo {author} {\bibfnamefont {A.}~\bibnamefont {Bansil}}, \bibinfo {author} {\bibfnamefont {S.}~\bibnamefont {Kaprzyk}}, \bibinfo {author} {\bibfnamefont {A.~T.}\ \bibnamefont {Stewart}}, \bibinfo {author} {\bibfnamefont {Y.}~\bibnamefont {Nagashima}}, \bibinfo {author} {\bibfnamefont {T.}~\bibnamefont {Hyodo}}, \bibinfo {author} {\bibfnamefont {S.}~\bibnamefont {Nanao}}, \bibinfo {author} {\bibfnamefont {H.}~\bibnamefont {Kawata}},\ and\ \bibinfo {author} {\bibfnamefont {N.}~\bibnamefont {Shiotani}},\ }\bibfield  {title} {\bibinfo {title} {High-resolution {C}ompton scattering study of {Li}: {A}sphericity of the {F}ermi surface and electron correlation effects},\ }\href {https://doi.org/10.1103/PhysRevLett.74.2252} {\bibfield  {journal} {\bibinfo  {journal} {Phys. Rev. Lett.}\ }\textbf {\bibinfo {volume} {74}},\ \bibinfo {pages} {2252} (\bibinfo {year}
  {1995}{\natexlab{a}})}\BibitemShut {NoStop}%
\bibitem [{\citenamefont {Sch\"ulke}\ \emph {et~al.}(2001)\citenamefont {Sch\"ulke}, \citenamefont {Sternemann}, \citenamefont {Kaprolat},\ and\ \citenamefont {D\"oring}}]{schulke:01}%
  \BibitemOpen
  \bibfield  {author} {\bibinfo {author} {\bibfnamefont {W.}~\bibnamefont {Sch\"ulke}}, \bibinfo {author} {\bibfnamefont {C.}~\bibnamefont {Sternemann}}, \bibinfo {author} {\bibfnamefont {A.}~\bibnamefont {Kaprolat}},\ and\ \bibinfo {author} {\bibfnamefont {G.}~\bibnamefont {D\"oring}},\ }\bibfield  {title} {\bibinfo {title} {Ultra-high resolution {C}ompton scattering of {Li} metal: {E}valuation with respect to the correlation corrected occupation number density},\ }\href {https://doi.org/https://doi.org/10.1524/zpch.2001.215.11.1353} {\bibfield  {journal} {\bibinfo  {journal} {Zeitschrift f\"ur Physikalische Chemie}\ }\textbf {\bibinfo {volume} {215}},\ \bibinfo {pages} {1353} (\bibinfo {year} {01 Nov. 2001})}\BibitemShut {NoStop}%
\bibitem [{\citenamefont {Erba}\ \emph {et~al.}(2011)\citenamefont {Erba}, \citenamefont {Itou}, \citenamefont {Sakurai}, \citenamefont {Yamaki}, \citenamefont {Ito}, \citenamefont {Casassa}, \citenamefont {Maschio}, \citenamefont {Terentjevs},\ and\ \citenamefont {Pisani}}]{Erba_2011}%
  \BibitemOpen
  \bibfield  {author} {\bibinfo {author} {\bibfnamefont {A.}~\bibnamefont {Erba}}, \bibinfo {author} {\bibfnamefont {M.}~\bibnamefont {Itou}}, \bibinfo {author} {\bibfnamefont {Y.}~\bibnamefont {Sakurai}}, \bibinfo {author} {\bibfnamefont {R.}~\bibnamefont {Yamaki}}, \bibinfo {author} {\bibfnamefont {M.}~\bibnamefont {Ito}}, \bibinfo {author} {\bibfnamefont {S.}~\bibnamefont {Casassa}}, \bibinfo {author} {\bibfnamefont {L.}~\bibnamefont {Maschio}}, \bibinfo {author} {\bibfnamefont {A.}~\bibnamefont {Terentjevs}},\ and\ \bibinfo {author} {\bibfnamefont {C.}~\bibnamefont {Pisani}},\ }\bibfield  {title} {\bibinfo {title} {{Beyond a single-determinantal description of the density matrix of periodic systems: Experimental versus theoretical Compton profiles of crystalline silicon}},\ }\href {https://doi.org/10.1103/PhysRevB.83.125208} {\bibfield  {journal} {\bibinfo  {journal} {Phys. Rev. B}\ }\textbf {\bibinfo {volume} {83}},\ \bibinfo {pages} {125208} (\bibinfo {year} {2011})}\BibitemShut {NoStop}%
\bibitem [{\citenamefont {Pisani}\ \emph {et~al.}(2011)\citenamefont {Pisani}, \citenamefont {Itou}, \citenamefont {Sakurai}, \citenamefont {Yamaki}, \citenamefont {Ito}, \citenamefont {Erba},\ and\ \citenamefont {Maschio}}]{Pisani_2011}%
  \BibitemOpen
  \bibfield  {author} {\bibinfo {author} {\bibfnamefont {C.}~\bibnamefont {Pisani}}, \bibinfo {author} {\bibfnamefont {M.}~\bibnamefont {Itou}}, \bibinfo {author} {\bibfnamefont {Y.}~\bibnamefont {Sakurai}}, \bibinfo {author} {\bibfnamefont {R.}~\bibnamefont {Yamaki}}, \bibinfo {author} {\bibfnamefont {M.}~\bibnamefont {Ito}}, \bibinfo {author} {\bibfnamefont {A.}~\bibnamefont {Erba}},\ and\ \bibinfo {author} {\bibfnamefont {L.}~\bibnamefont {Maschio}},\ }\bibfield  {title} {\bibinfo {title} {{"Evidence of instantaneous electron correlation from Compton profiles of crystalline silicon"}},\ }\href {https://doi.org/10.1039/C0CP01604G} {\bibfield  {journal} {\bibinfo  {journal} {Phys. Chem. Chem. Phys.}\ }\textbf {\bibinfo {volume} {13}},\ \bibinfo {pages} {933} (\bibinfo {year} {2011})}\BibitemShut {NoStop}%
\bibitem [{\citenamefont {Dugdale}\ \emph {et~al.}(2000)\citenamefont {Dugdale}, \citenamefont {Fretwell}, \citenamefont {Chen}, \citenamefont {Tanaka}, \citenamefont {Shukla}, \citenamefont {Buslaps}, \citenamefont {Bellin}, \citenamefont {Loupias}, \citenamefont {Alam}, \citenamefont {Manuel}, \citenamefont {Suortti},\ and\ \citenamefont {Shiotani}}]{DUGDALE2000361}%
  \BibitemOpen
  \bibfield  {author} {\bibinfo {author} {\bibfnamefont {S.~B.}\ \bibnamefont {Dugdale}}, \bibinfo {author} {\bibfnamefont {H.}~\bibnamefont {Fretwell}}, \bibinfo {author} {\bibfnamefont {K.~J.}\ \bibnamefont {Chen}}, \bibinfo {author} {\bibfnamefont {Y.}~\bibnamefont {Tanaka}}, \bibinfo {author} {\bibfnamefont {A.}~\bibnamefont {Shukla}}, \bibinfo {author} {\bibfnamefont {T.}~\bibnamefont {Buslaps}}, \bibinfo {author} {\bibfnamefont {C.}~\bibnamefont {Bellin}}, \bibinfo {author} {\bibfnamefont {G.}~\bibnamefont {Loupias}}, \bibinfo {author} {\bibfnamefont {M.~A.}\ \bibnamefont {Alam}}, \bibinfo {author} {\bibfnamefont {A.~A.}\ \bibnamefont {Manuel}}, \bibinfo {author} {\bibfnamefont {P.}~\bibnamefont {Suortti}},\ and\ \bibinfo {author} {\bibfnamefont {N.}~\bibnamefont {Shiotani}},\ }\bibfield  {title} {\bibinfo {title} {{A high-resolution Compton scattering study of the Fermi surface of chromium}},\ }\href {https://doi.org/https://doi.org/10.1016/S0022-3697(99)00317-0} {\bibfield  {journal} {\bibinfo
  {journal} {Journal of Physics and Chemistry of Solids}\ }\textbf {\bibinfo {volume} {61}},\ \bibinfo {pages} {361} (\bibinfo {year} {2000})}\BibitemShut {NoStop}%
\bibitem [{\citenamefont {Dixon}\ \emph {et~al.}(1998)\citenamefont {Dixon}, \citenamefont {Duffy}, \citenamefont {Gardelis}, \citenamefont {McCarthy}, \citenamefont {Cooper}, \citenamefont {Dugdale}, \citenamefont {Jarlborg},\ and\ \citenamefont {Timms}}]{di.du.98}%
  \BibitemOpen
  \bibfield  {author} {\bibinfo {author} {\bibfnamefont {M.~A.~G.}\ \bibnamefont {Dixon}}, \bibinfo {author} {\bibfnamefont {J.~A.}\ \bibnamefont {Duffy}}, \bibinfo {author} {\bibfnamefont {S.}~\bibnamefont {Gardelis}}, \bibinfo {author} {\bibfnamefont {J.~E.}\ \bibnamefont {McCarthy}}, \bibinfo {author} {\bibfnamefont {M.~J.}\ \bibnamefont {Cooper}}, \bibinfo {author} {\bibfnamefont {S.~B.}\ \bibnamefont {Dugdale}}, \bibinfo {author} {\bibfnamefont {T.}~\bibnamefont {Jarlborg}},\ and\ \bibinfo {author} {\bibfnamefont {D.~N.}\ \bibnamefont {Timms}},\ }\bibfield  {title} {\bibinfo {title} {Spin density in ferromagnetic nickel: {A} magnetic {C}ompton scattering study},\ }\href {https://doi.org/10.1088/0953-8984/10/12/014} {\bibfield  {journal} {\bibinfo  {journal} {Journal of Physics: Condensed Matter}\ }\textbf {\bibinfo {volume} {10}},\ \bibinfo {pages} {2759} (\bibinfo {year} {1998})}\BibitemShut {NoStop}%
\bibitem [{\citenamefont {Mandal}\ \emph {et~al.}(2022)\citenamefont {Mandal}, \citenamefont {Haule}, \citenamefont {Rabe},\ and\ \citenamefont {Vanderbilt}}]{mandal2022electronic}%
  \BibitemOpen
  \bibfield  {author} {\bibinfo {author} {\bibfnamefont {S.}~\bibnamefont {Mandal}}, \bibinfo {author} {\bibfnamefont {K.}~\bibnamefont {Haule}}, \bibinfo {author} {\bibfnamefont {K.~M.}\ \bibnamefont {Rabe}},\ and\ \bibinfo {author} {\bibfnamefont {D.}~\bibnamefont {Vanderbilt}},\ }\bibfield  {title} {\bibinfo {title} {{Electronic correlation in nearly free electron metals with beyond-DFT methods}},\ }\href {https://doi.org/https://doi.org/10.1038/s41524-022-00867-8} {\bibfield  {journal} {\bibinfo  {journal} {npj Computational Materials}\ }\textbf {\bibinfo {volume} {8}},\ \bibinfo {pages} {181} (\bibinfo {year} {2022})}\BibitemShut {NoStop}%
\bibitem [{\citenamefont {Friedrich}\ \emph {et~al.}(2022)\citenamefont {Friedrich}, \citenamefont {Bl{\"u}gel},\ and\ \citenamefont {Nabok}}]{Friedrich-QSGW}%
  \BibitemOpen
  \bibfield  {author} {\bibinfo {author} {\bibfnamefont {C.}~\bibnamefont {Friedrich}}, \bibinfo {author} {\bibfnamefont {S.}~\bibnamefont {Bl{\"u}gel}},\ and\ \bibinfo {author} {\bibfnamefont {D.}~\bibnamefont {Nabok}},\ }\bibfield  {title} {\bibinfo {title} {{Quasiparticle self-consistent GW study of simple metals}},\ }\href {https://doi.org/10.3390/nano12203660} {\bibfield  {journal} {\bibinfo  {journal} {Nanomaterials}\ }\textbf {\bibinfo {volume} {12}},\ \bibinfo {pages} {3660} (\bibinfo {year} {2022})}\BibitemShut {NoStop}%
\bibitem [{\citenamefont {Hedin}(1965)}]{hedin_new_1965}%
  \BibitemOpen
  \bibfield  {author} {\bibinfo {author} {\bibfnamefont {L.}~\bibnamefont {Hedin}},\ }\bibfield  {title} {\bibinfo {title} {New {Method} for {Calculating} the {One}-{Particle} {Green}'s {Function} with {Application} to the {Electron}-{Gas} {Problem}},\ }\href {https://doi.org/10.1103/PhysRev.139.A796} {\bibfield  {journal} {\bibinfo  {journal} {Phys. Rev.}\ }\textbf {\bibinfo {volume} {139}},\ \bibinfo {pages} {A796} (\bibinfo {year} {1965})}\BibitemShut {NoStop}%
\bibitem [{\citenamefont {Pashov}\ \emph {et~al.}(2020)\citenamefont {Pashov}, \citenamefont {Acharya}, \citenamefont {Lambrecht}, \citenamefont {Jackson}, \citenamefont {Belashchenko}, \citenamefont {Chantis}, \citenamefont {Jamet},\ and\ \citenamefont {Van~Schilfgaarde}}]{questaal-cpc}%
  \BibitemOpen
  \bibfield  {author} {\bibinfo {author} {\bibfnamefont {D.}~\bibnamefont {Pashov}}, \bibinfo {author} {\bibfnamefont {S.}~\bibnamefont {Acharya}}, \bibinfo {author} {\bibfnamefont {W.~R.~L.}\ \bibnamefont {Lambrecht}}, \bibinfo {author} {\bibfnamefont {J.}~\bibnamefont {Jackson}}, \bibinfo {author} {\bibfnamefont {K.~D.}\ \bibnamefont {Belashchenko}}, \bibinfo {author} {\bibfnamefont {A.}~\bibnamefont {Chantis}}, \bibinfo {author} {\bibfnamefont {F.}~\bibnamefont {Jamet}},\ and\ \bibinfo {author} {\bibfnamefont {M.}~\bibnamefont {Van~Schilfgaarde}},\ }\bibfield  {title} {\bibinfo {title} {{Questaal: {A} package of electronic structure methods based on the linear muffin-tin orbital technique}},\ }\href {https://doi.org/10.1016/j.cpc.2019.107065} {\bibfield  {journal} {\bibinfo  {journal} {Computer Physics Communications}\ }\textbf {\bibinfo {volume} {249}},\ \bibinfo {pages} {107065} (\bibinfo {year} {2020})}\BibitemShut {NoStop}%
\bibitem [{que(2026)}]{questaal-website}%
  \BibitemOpen
  \href {https://www.questaal.org} {\bibinfo {title} {{Questaal website}}},\ \bibinfo {howpublished} {\url{https://www.questaal.org}} (\bibinfo {year} {2026})\BibitemShut {NoStop}%
\bibitem [{\citenamefont {Dewhurst}\ \emph {et~al.}()\citenamefont {Dewhurst}, \citenamefont {Sharma}, \citenamefont {Nordstr\"{o}m}, \citenamefont {Cricchio}, \citenamefont {Granas},\ and\ \citenamefont {Gross}}]{elk}%
  \BibitemOpen
  \bibfield  {author} {\bibinfo {author} {\bibfnamefont {J.~K.}\ \bibnamefont {Dewhurst}}, \bibinfo {author} {\bibfnamefont {S.}~\bibnamefont {Sharma}}, \bibinfo {author} {\bibfnamefont {L.}~\bibnamefont {Nordstr\"{o}m}}, \bibinfo {author} {\bibfnamefont {F.}~\bibnamefont {Cricchio}}, \bibinfo {author} {\bibfnamefont {O.}~\bibnamefont {Granas}},\ and\ \bibinfo {author} {\bibfnamefont {E.~K.~U.}\ \bibnamefont {Gross}},\ }\href@noop {} {\bibinfo {title} {{The Elk Code}}},\ \bibinfo {howpublished} {\url{http://elk.sourceforge.net/}}\BibitemShut {NoStop}%
\bibitem [{\citenamefont {Ernsting}\ \emph {et~al.}(2014)\citenamefont {Ernsting}, \citenamefont {Billington}, \citenamefont {Haynes}, \citenamefont {Millichamp}, \citenamefont {Taylor}, \citenamefont {Duffy}, \citenamefont {Giblin}, \citenamefont {Dewhurst},\ and\ \citenamefont {Dugdale}}]{Ernsting_2014}%
  \BibitemOpen
  \bibfield  {author} {\bibinfo {author} {\bibfnamefont {D.}~\bibnamefont {Ernsting}}, \bibinfo {author} {\bibfnamefont {D.}~\bibnamefont {Billington}}, \bibinfo {author} {\bibfnamefont {T.~D.}\ \bibnamefont {Haynes}}, \bibinfo {author} {\bibfnamefont {T.~E.}\ \bibnamefont {Millichamp}}, \bibinfo {author} {\bibfnamefont {J.~W.}\ \bibnamefont {Taylor}}, \bibinfo {author} {\bibfnamefont {J.~A.}\ \bibnamefont {Duffy}}, \bibinfo {author} {\bibfnamefont {S.~R.}\ \bibnamefont {Giblin}}, \bibinfo {author} {\bibfnamefont {J.~K.}\ \bibnamefont {Dewhurst}},\ and\ \bibinfo {author} {\bibfnamefont {S.~B.}\ \bibnamefont {Dugdale}},\ }\bibfield  {title} {\bibinfo {title} {Calculating electron momentum densities and {C}ompton profiles using the linear tetrahedron method},\ }\href {https://doi.org/10.1088/0953-8984/26/49/495501} {\bibfield  {journal} {\bibinfo  {journal} {Journal of Physics: Condensed Matter}\ }\textbf {\bibinfo {volume} {26}},\ \bibinfo {pages} {495501} (\bibinfo {year} {2014})}\BibitemShut {NoStop}%
\bibitem [{\citenamefont {Golze}\ \emph {et~al.}(2019)\citenamefont {Golze}, \citenamefont {Dvorak},\ and\ \citenamefont {Rinke}}]{gw_compendium}%
  \BibitemOpen
  \bibfield  {author} {\bibinfo {author} {\bibfnamefont {D.}~\bibnamefont {Golze}}, \bibinfo {author} {\bibfnamefont {M.}~\bibnamefont {Dvorak}},\ and\ \bibinfo {author} {\bibfnamefont {P.}~\bibnamefont {Rinke}},\ }\bibfield  {title} {\bibinfo {title} {{The {GW} {Compendium}: {A} {Practical} {Guide} to {Theoretical} {Photoemission} {Spectroscopy}}},\ }\href {https://doi.org/10.3389/fchem.2019.00377} {\bibfield  {journal} {\bibinfo  {journal} {Frontiers in Chemistry}\ }\textbf {\bibinfo {volume} {7}},\ \bibinfo {pages} {377} (\bibinfo {year} {2019})}\BibitemShut {NoStop}%
\bibitem [{\citenamefont {Kotani}\ \emph {et~al.}(2007{\natexlab{a}})\citenamefont {Kotani}, \citenamefont {Van~Schilfgaarde},\ and\ \citenamefont {Faleev}}]{qsgw-original}%
  \BibitemOpen
  \bibfield  {author} {\bibinfo {author} {\bibfnamefont {T.}~\bibnamefont {Kotani}}, \bibinfo {author} {\bibfnamefont {M.}~\bibnamefont {Van~Schilfgaarde}},\ and\ \bibinfo {author} {\bibfnamefont {S.~V.}\ \bibnamefont {Faleev}},\ }\bibfield  {title} {\bibinfo {title} {{Quasiparticle self-consistent {GW} method: {A} basis for the independent-particle approximation}},\ }\href {https://doi.org/10.1103/PhysRevB.76.165106} {\bibfield  {journal} {\bibinfo  {journal} {Physical Review B}\ }\textbf {\bibinfo {volume} {76}},\ \bibinfo {pages} {165106} (\bibinfo {year} {2007}{\natexlab{a}})}\BibitemShut {NoStop}%
\bibitem [{\citenamefont {Cooper}(1985)}]{cooper:85}%
  \BibitemOpen
  \bibfield  {author} {\bibinfo {author} {\bibfnamefont {M.~J.}\ \bibnamefont {Cooper}},\ }\bibfield  {title} {\bibinfo {title} {{C}ompton scattering and electron momentum determination},\ }\href {https://doi.org/10.1088/0034-4885/48/4/001} {\bibfield  {journal} {\bibinfo  {journal} {Reports on Progress in Physics}\ }\textbf {\bibinfo {volume} {48}},\ \bibinfo {pages} {415} (\bibinfo {year} {1985})}\BibitemShut {NoStop}%
\bibitem [{\citenamefont {Cooper}(1997)}]{cooper_compton_1997}%
  \BibitemOpen
  \bibfield  {author} {\bibinfo {author} {\bibfnamefont {M.~J.}\ \bibnamefont {Cooper}},\ }\bibfield  {title} {\bibinfo {title} {Compton scattering and the study of electron momentum density distributions},\ }\href {https://doi.org/10.1016/S0969-806X(97)00024-8} {\bibfield  {journal} {\bibinfo  {journal} {Radiation Physics and Chemistry}\ }\bibinfo {series} {Inelastic {Scattering} of {X}-{Rays} and {Gamma} {Rays}},\ \textbf {\bibinfo {volume} {50}},\ \bibinfo {pages} {63} (\bibinfo {year} {1997})}\BibitemShut {NoStop}%
\bibitem [{\citenamefont {Sakai}(1996)}]{sakai_magnetic_1996}%
  \BibitemOpen
  \bibfield  {author} {\bibinfo {author} {\bibfnamefont {N.}~\bibnamefont {Sakai}},\ }\bibfield  {title} {\bibinfo {title} {Magnetic {Compton} {Scattering} and {Measurements} of {Momentum} {Distribution} of {Magnetic} {Electrons}},\ }\href {https://doi.org/10.1107/S0021889895007126} {\bibfield  {journal} {\bibinfo  {journal} {J. Appl. Crystallogr.}\ }\textbf {\bibinfo {volume} {29}},\ \bibinfo {pages} {81} (\bibinfo {year} {1996})}\BibitemShut {NoStop}%
\bibitem [{\citenamefont {Kontrym-Sznajd}(2009)}]{kontrym-sznajd_fermiology_2009}%
  \BibitemOpen
  \bibfield  {author} {\bibinfo {author} {\bibfnamefont {G.}~\bibnamefont {Kontrym-Sznajd}},\ }\bibfield  {title} {\bibinfo {title} {Fermiology via the electron momentum distribution ({Review} {Article})},\ }\href {https://doi.org/10.1063/1.3224712} {\bibfield  {journal} {\bibinfo  {journal} {Low Temperature Physics}\ }\textbf {\bibinfo {volume} {35}},\ \bibinfo {pages} {599} (\bibinfo {year} {2009})}\BibitemShut {NoStop}%
\bibitem [{\citenamefont {Ketels}\ \emph {et~al.}(2021)\citenamefont {Ketels}, \citenamefont {Billington}, \citenamefont {Dugdale}, \citenamefont {Leitner},\ and\ \citenamefont {Hugenschmidt}}]{ketels_momentum_2021}%
  \BibitemOpen
  \bibfield  {author} {\bibinfo {author} {\bibfnamefont {J.}~\bibnamefont {Ketels}}, \bibinfo {author} {\bibfnamefont {D.}~\bibnamefont {Billington}}, \bibinfo {author} {\bibfnamefont {S.~B.}\ \bibnamefont {Dugdale}}, \bibinfo {author} {\bibfnamefont {M.}~\bibnamefont {Leitner}},\ and\ \bibinfo {author} {\bibfnamefont {C.~P.}\ \bibnamefont {Hugenschmidt}},\ }\bibfield  {title} {\bibinfo {title} {Momentum density spectroscopy of {Pd}: {Comparison} of {2D}-{ACAR} and {Compton} scattering using a {1D}-to-{2D} reconstruction method},\ }\href {https://doi.org/10.1103/PhysRevB.104.075160} {\bibfield  {journal} {\bibinfo  {journal} {Phys. Rev. B}\ }\textbf {\bibinfo {volume} {104}},\ \bibinfo {pages} {075160} (\bibinfo {year} {2021})}\BibitemShut {NoStop}%
\bibitem [{\citenamefont {Olevano}\ \emph {et~al.}(2012)\citenamefont {Olevano}, \citenamefont {Titov}, \citenamefont {Ladisa}, \citenamefont {Hämäläinen}, \citenamefont {Huotari},\ and\ \citenamefont {Holzmann}}]{olevano_momentum_2012}%
  \BibitemOpen
  \bibfield  {author} {\bibinfo {author} {\bibfnamefont {V.}~\bibnamefont {Olevano}}, \bibinfo {author} {\bibfnamefont {A.}~\bibnamefont {Titov}}, \bibinfo {author} {\bibfnamefont {M.}~\bibnamefont {Ladisa}}, \bibinfo {author} {\bibfnamefont {K.}~\bibnamefont {Hämäläinen}}, \bibinfo {author} {\bibfnamefont {S.}~\bibnamefont {Huotari}},\ and\ \bibinfo {author} {\bibfnamefont {M.}~\bibnamefont {Holzmann}},\ }\bibfield  {title} {\bibinfo {title} {Momentum distribution and {Compton} profile by the ab initio {GW} approximation},\ }\href {https://doi.org/10.1103/PhysRevB.86.195123} {\bibfield  {journal} {\bibinfo  {journal} {Phys. Rev. B}\ }\textbf {\bibinfo {volume} {86}},\ \bibinfo {pages} {195123} (\bibinfo {year} {2012})}\BibitemShut {NoStop}%
\bibitem [{\citenamefont {Barbiellini}\ and\ \citenamefont {Bansil}(2001)}]{barbiellini2001EMD}%
  \BibitemOpen
  \bibfield  {author} {\bibinfo {author} {\bibfnamefont {B.}~\bibnamefont {Barbiellini}}\ and\ \bibinfo {author} {\bibfnamefont {A.}~\bibnamefont {Bansil}},\ }\bibfield  {title} {\bibinfo {title} {Treatment of correlation effects in electron momentum density: density functional theory and beyond},\ }\href {https://doi.org/https://doi.org/10.1016/S0022-3697(01)00176-7} {\bibfield  {journal} {\bibinfo  {journal} {Journal of Physics and Chemistry of Solids}\ }\textbf {\bibinfo {volume} {62}},\ \bibinfo {pages} {2181} (\bibinfo {year} {2001})}\BibitemShut {NoStop}%
\bibitem [{\citenamefont {James}\ \emph {et~al.}(2021)\citenamefont {James}, \citenamefont {Sekania}, \citenamefont {Dugdale},\ and\ \citenamefont {Chioncel}}]{james2021magnetic}%
  \BibitemOpen
  \bibfield  {author} {\bibinfo {author} {\bibfnamefont {A.~D.~N.}\ \bibnamefont {James}}, \bibinfo {author} {\bibfnamefont {M.}~\bibnamefont {Sekania}}, \bibinfo {author} {\bibfnamefont {S.~B.}\ \bibnamefont {Dugdale}},\ and\ \bibinfo {author} {\bibfnamefont {L.}~\bibnamefont {Chioncel}},\ }\bibfield  {title} {\bibinfo {title} {{Magnetic Compton profiles of Ni beyond the one-particle picture: Numerically exact and perturbative solvers of dynamical mean-field theory}},\ }\href {https://doi.org/10.1103/PhysRevB.103.115144} {\bibfield  {journal} {\bibinfo  {journal} {Phys. Rev. B}\ }\textbf {\bibinfo {volume} {103}},\ \bibinfo {pages} {115144} (\bibinfo {year} {2021})}\BibitemShut {NoStop}%
\bibitem [{\citenamefont {Lam}\ and\ \citenamefont {Platzman}(1974{\natexlab{a}})}]{lam_momentum_1974_I}%
  \BibitemOpen
  \bibfield  {author} {\bibinfo {author} {\bibfnamefont {L.}~\bibnamefont {Lam}}\ and\ \bibinfo {author} {\bibfnamefont {P.~M.}\ \bibnamefont {Platzman}},\ }\bibfield  {title} {\bibinfo {title} {{Momentum density and Compton profile of the inhomogeneous interacting electronic system. I. Formalism}},\ }\href {https://doi.org/10.1103/PhysRevB.9.5122} {\bibfield  {journal} {\bibinfo  {journal} {Phys. Rev. B}\ }\textbf {\bibinfo {volume} {9}},\ \bibinfo {pages} {5122} (\bibinfo {year} {1974}{\natexlab{a}})}\BibitemShut {NoStop}%
\bibitem [{\citenamefont {Lam}\ and\ \citenamefont {Platzman}(1974{\natexlab{b}})}]{lam_momentum_1974}%
  \BibitemOpen
  \bibfield  {author} {\bibinfo {author} {\bibfnamefont {L.}~\bibnamefont {Lam}}\ and\ \bibinfo {author} {\bibfnamefont {P.~M.}\ \bibnamefont {Platzman}},\ }\bibfield  {title} {\bibinfo {title} {{Momentum density and {Compton} profile of the inhomogeneous interacting electron system. {II}. {Application} to atoms}},\ }\href {https://doi.org/10.1103/PhysRevB.9.5128} {\bibfield  {journal} {\bibinfo  {journal} {Phys. Rev. B}\ }\textbf {\bibinfo {volume} {9}},\ \bibinfo {pages} {5128} (\bibinfo {year} {1974}{\natexlab{b}})}\BibitemShut {NoStop}%
\bibitem [{\citenamefont {Sakurai}\ \emph {et~al.}(1995{\natexlab{b}})\citenamefont {Sakurai}, \citenamefont {Tanaka}, \citenamefont {Bansil}, \citenamefont {Kaprzyk}, \citenamefont {Stewart}, \citenamefont {Nagashima}, \citenamefont {Hyodo}, \citenamefont {Nanao}, \citenamefont {Kawata},\ and\ \citenamefont {Shiotani}}]{Sakurai-Li}%
  \BibitemOpen
  \bibfield  {author} {\bibinfo {author} {\bibfnamefont {Y.}~\bibnamefont {Sakurai}}, \bibinfo {author} {\bibfnamefont {Y.}~\bibnamefont {Tanaka}}, \bibinfo {author} {\bibfnamefont {A.}~\bibnamefont {Bansil}}, \bibinfo {author} {\bibfnamefont {S.}~\bibnamefont {Kaprzyk}}, \bibinfo {author} {\bibfnamefont {A.~T.}\ \bibnamefont {Stewart}}, \bibinfo {author} {\bibfnamefont {Y.}~\bibnamefont {Nagashima}}, \bibinfo {author} {\bibfnamefont {T.}~\bibnamefont {Hyodo}}, \bibinfo {author} {\bibfnamefont {S.}~\bibnamefont {Nanao}}, \bibinfo {author} {\bibfnamefont {H.}~\bibnamefont {Kawata}},\ and\ \bibinfo {author} {\bibfnamefont {N.}~\bibnamefont {Shiotani}},\ }\bibfield  {title} {\bibinfo {title} {High-resolution compton scattering study of li: Asphericity of the fermi surface and electron correlation effects},\ }\href {https://doi.org/10.1103/PhysRevLett.74.2252} {\bibfield  {journal} {\bibinfo  {journal} {Phys. Rev. Lett.}\ }\textbf {\bibinfo {volume} {74}},\ \bibinfo {pages} {2252} (\bibinfo {year}
  {1995}{\natexlab{b}})}\BibitemShut {NoStop}%
\bibitem [{\citenamefont {Sch\"ulke}\ \emph {et~al.}(1996)\citenamefont {Sch\"ulke}, \citenamefont {Stutz}, \citenamefont {Wohlert},\ and\ \citenamefont {Kaprolat}}]{Schulke-Li}%
  \BibitemOpen
  \bibfield  {author} {\bibinfo {author} {\bibfnamefont {W.}~\bibnamefont {Sch\"ulke}}, \bibinfo {author} {\bibfnamefont {G.}~\bibnamefont {Stutz}}, \bibinfo {author} {\bibfnamefont {F.}~\bibnamefont {Wohlert}},\ and\ \bibinfo {author} {\bibfnamefont {A.}~\bibnamefont {Kaprolat}},\ }\bibfield  {title} {\bibinfo {title} {{Electron momentum-space densities of Li metal: A high-resolution Compton-scattering study}},\ }\href {https://doi.org/10.1103/PhysRevB.54.14381} {\bibfield  {journal} {\bibinfo  {journal} {Phys. Rev. B}\ }\textbf {\bibinfo {volume} {54}},\ \bibinfo {pages} {14381} (\bibinfo {year} {1996})}\BibitemShut {NoStop}%
\bibitem [{\citenamefont {Perdew}\ and\ \citenamefont {Wang}(1992)}]{perdew1992}%
  \BibitemOpen
  \bibfield  {author} {\bibinfo {author} {\bibfnamefont {J.}~\bibnamefont {Perdew}}\ and\ \bibinfo {author} {\bibfnamefont {Y.}~\bibnamefont {Wang}},\ }\bibfield  {title} {\bibinfo {title} {Accurate and simple analytic representation of the electron-gas correlation energy},\ }\href {https://doi.org/10.1103/PhysRevB.45.13244} {\bibfield  {journal} {\bibinfo  {journal} {Phys. Rev. B}\ }\textbf {\bibinfo {volume} {45}},\ \bibinfo {pages} {13244} (\bibinfo {year} {1992})}\BibitemShut {NoStop}%
\bibitem [{\citenamefont {James}\ \emph {et~al.}(2025)\citenamefont {James}, \citenamefont {Harris-Lee},\ and\ \citenamefont {Dugdale}}]{James_2025}%
  \BibitemOpen
  \bibfield  {author} {\bibinfo {author} {\bibfnamefont {A.~D.~N.}\ \bibnamefont {James}}, \bibinfo {author} {\bibfnamefont {E.~I.}\ \bibnamefont {Harris-Lee}},\ and\ \bibinfo {author} {\bibfnamefont {S.~B.}\ \bibnamefont {Dugdale}},\ }\bibfield  {title} {\bibinfo {title} {{Investigating the magnetism of Ni from a momentum space perspective}},\ }\href {https://doi.org/10.1088/2516-1075/add9f6} {\bibfield  {journal} {\bibinfo  {journal} {Electronic Structure}\ }\textbf {\bibinfo {volume} {7}},\ \bibinfo {pages} {025005} (\bibinfo {year} {2025})}\BibitemShut {NoStop}%
\bibitem [{dat(2026)}]{data-archive}%
  \BibitemOpen
  \href {https://www.dropbox.com/scl/fo/ys9yushvkzjsgkgonpfru/ABbr1hIO1db4xgIxMwsym0Q?rlkey=vlhgx4n6t7x39cbsa4cff15o7&st=wv5psd0q&dl=0} {\bibinfo {title} {{Data Archive URL}}},\ \bibinfo {howpublished} {\url{https://www.dropbox.com/scl/fo/ys9yushvkzjsgkgonpfru/ABbr1hIO1db4xgIxMwsym0Q?rlkey=vlhgx4n6t7x39cbsa4cff15o7&st=wv5psd0q&dl=0}} (\bibinfo {year} {2026})\BibitemShut {NoStop}%
\bibitem [{\citenamefont {Hiraoka}\ \emph {et~al.}(2020)\citenamefont {Hiraoka}, \citenamefont {Yang}, \citenamefont {Hagiya}, \citenamefont {Niozu}, \citenamefont {Matsuda}, \citenamefont {Huotari}, \citenamefont {Holzmann},\ and\ \citenamefont {Ceperley}}]{Hiraoka-Li-Z-Compton}%
  \BibitemOpen
  \bibfield  {author} {\bibinfo {author} {\bibfnamefont {N.}~\bibnamefont {Hiraoka}}, \bibinfo {author} {\bibfnamefont {Y.}~\bibnamefont {Yang}}, \bibinfo {author} {\bibfnamefont {T.}~\bibnamefont {Hagiya}}, \bibinfo {author} {\bibfnamefont {A.}~\bibnamefont {Niozu}}, \bibinfo {author} {\bibfnamefont {K.}~\bibnamefont {Matsuda}}, \bibinfo {author} {\bibfnamefont {S.}~\bibnamefont {Huotari}}, \bibinfo {author} {\bibfnamefont {M.}~\bibnamefont {Holzmann}},\ and\ \bibinfo {author} {\bibfnamefont {D.~M.}\ \bibnamefont {Ceperley}},\ }\bibfield  {title} {\bibinfo {title} {Direct observation of the momentum distribution and renormalization factor in lithium},\ }\href {https://doi.org/10.1103/PhysRevB.101.165124} {\bibfield  {journal} {\bibinfo  {journal} {Phys. Rev. B}\ }\textbf {\bibinfo {volume} {101}},\ \bibinfo {pages} {165124} (\bibinfo {year} {2020})}\BibitemShut {NoStop}%
\bibitem [{\citenamefont {Sternemann}\ \emph {et~al.}(2001)\citenamefont {Sternemann}, \citenamefont {Buslaps}, \citenamefont {Shukla}, \citenamefont {Suortti}, \citenamefont {D\"oring},\ and\ \citenamefont {Sch\"ulke}}]{Sternemann-Li-temp-CPs}%
  \BibitemOpen
  \bibfield  {author} {\bibinfo {author} {\bibfnamefont {C.}~\bibnamefont {Sternemann}}, \bibinfo {author} {\bibfnamefont {T.}~\bibnamefont {Buslaps}}, \bibinfo {author} {\bibfnamefont {A.}~\bibnamefont {Shukla}}, \bibinfo {author} {\bibfnamefont {P.}~\bibnamefont {Suortti}}, \bibinfo {author} {\bibfnamefont {G.}~\bibnamefont {D\"oring}},\ and\ \bibinfo {author} {\bibfnamefont {W.}~\bibnamefont {Sch\"ulke}},\ }\bibfield  {title} {\bibinfo {title} {{Temperature influence on the valence Compton profiles of aluminum and lithium}},\ }\href {https://doi.org/10.1103/PhysRevB.63.094301} {\bibfield  {journal} {\bibinfo  {journal} {Phys. Rev. B}\ }\textbf {\bibinfo {volume} {63}},\ \bibinfo {pages} {094301} (\bibinfo {year} {2001})}\BibitemShut {NoStop}%
\bibitem [{\citenamefont {Chen}\ \emph {et~al.}(1999)\citenamefont {Chen}, \citenamefont {Caspar}, \citenamefont {Bellin},\ and\ \citenamefont {Loupias}}]{Chen-Li-temp-CPs}%
  \BibitemOpen
  \bibfield  {author} {\bibinfo {author} {\bibfnamefont {K.~J.}\ \bibnamefont {Chen}}, \bibinfo {author} {\bibfnamefont {V.}~\bibnamefont {Caspar}}, \bibinfo {author} {\bibfnamefont {C.}~\bibnamefont {Bellin}},\ and\ \bibinfo {author} {\bibfnamefont {G.}~\bibnamefont {Loupias}},\ }\bibfield  {title} {\bibinfo {title} {{Investigation of temperature dependence of Compton profiles in lithium}},\ }\href {https://doi.org/https://doi.org/10.1016/S0038-1098(99)00094-0} {\bibfield  {journal} {\bibinfo  {journal} {Solid State Communications}\ }\textbf {\bibinfo {volume} {110}},\ \bibinfo {pages} {357} (\bibinfo {year} {1999})}\BibitemShut {NoStop}%
\bibitem [{\citenamefont {Kubo}(1996)}]{Kubo-Li-GW}%
  \BibitemOpen
  \bibfield  {author} {\bibinfo {author} {\bibfnamefont {Y.}~\bibnamefont {Kubo}},\ }\bibfield  {title} {\bibinfo {title} {{Compton Profiles of Li in the GW Approximation}},\ }\href {https://doi.org/10.1143/JPSJ.65.16} {\bibfield  {journal} {\bibinfo  {journal} {Journal of the Physical Society of Japan}\ }\textbf {\bibinfo {volume} {65}},\ \bibinfo {pages} {16} (\bibinfo {year} {1996})}\BibitemShut {NoStop}%
\bibitem [{\citenamefont {Vos}\ \emph {et~al.}(2002)\citenamefont {Vos}, \citenamefont {Kheifets}, \citenamefont {Sashin}, \citenamefont {Weigold}, \citenamefont {Usuda},\ and\ \citenamefont {Aryasetiawan}}]{Vos-Li-exp-Akw}%
  \BibitemOpen
  \bibfield  {author} {\bibinfo {author} {\bibfnamefont {M.}~\bibnamefont {Vos}}, \bibinfo {author} {\bibfnamefont {A.~S.}\ \bibnamefont {Kheifets}}, \bibinfo {author} {\bibfnamefont {V.~A.}\ \bibnamefont {Sashin}}, \bibinfo {author} {\bibfnamefont {E.}~\bibnamefont {Weigold}}, \bibinfo {author} {\bibfnamefont {M.}~\bibnamefont {Usuda}},\ and\ \bibinfo {author} {\bibfnamefont {F.}~\bibnamefont {Aryasetiawan}},\ }\bibfield  {title} {\bibinfo {title} {{Quantitative measurement of the spectral function of aluminum and lithium by electron momentum spectroscopy}},\ }\href {https://doi.org/10.1103/PhysRevB.66.155414} {\bibfield  {journal} {\bibinfo  {journal} {Phys. Rev. B}\ }\textbf {\bibinfo {volume} {66}},\ \bibinfo {pages} {155414} (\bibinfo {year} {2002})}\BibitemShut {NoStop}%
\bibitem [{\citenamefont {Dugdale}\ and\ \citenamefont {Jarlborg}(1998)}]{DUGDALE1998283}%
  \BibitemOpen
  \bibfield  {author} {\bibinfo {author} {\bibfnamefont {S.~B.}\ \bibnamefont {Dugdale}}\ and\ \bibinfo {author} {\bibfnamefont {T.}~\bibnamefont {Jarlborg}},\ }\bibfield  {title} {\bibinfo {title} {{Thermal disorder versus correlation in Compton profiles from alkali metals}},\ }\href {https://doi.org/https://doi.org/10.1016/S0038-1098(97)10112-0} {\bibfield  {journal} {\bibinfo  {journal} {Solid State Communications}\ }\textbf {\bibinfo {volume} {105}},\ \bibinfo {pages} {283} (\bibinfo {year} {1998})}\BibitemShut {NoStop}%
\bibitem [{\citenamefont {Filippi}\ and\ \citenamefont {Ceperley}(1999)}]{Filippi-Li-QMC}%
  \BibitemOpen
  \bibfield  {author} {\bibinfo {author} {\bibfnamefont {C.}~\bibnamefont {Filippi}}\ and\ \bibinfo {author} {\bibfnamefont {D.~M.}\ \bibnamefont {Ceperley}},\ }\bibfield  {title} {\bibinfo {title} {{Quantum Monte Carlo calculation of Compton profiles of solid lithium}},\ }\href {https://doi.org/10.1103/PhysRevB.59.7907} {\bibfield  {journal} {\bibinfo  {journal} {Phys. Rev. B}\ }\textbf {\bibinfo {volume} {59}},\ \bibinfo {pages} {7907} (\bibinfo {year} {1999})}\BibitemShut {NoStop}%
\bibitem [{\citenamefont {Yang}\ \emph {et~al.}(2020)\citenamefont {Yang}, \citenamefont {Hiraoka}, \citenamefont {Matsuda}, \citenamefont {Holzmann},\ and\ \citenamefont {Ceperley}}]{Yang-QMC-Li}%
  \BibitemOpen
  \bibfield  {author} {\bibinfo {author} {\bibfnamefont {Y.}~\bibnamefont {Yang}}, \bibinfo {author} {\bibfnamefont {N.}~\bibnamefont {Hiraoka}}, \bibinfo {author} {\bibfnamefont {K.}~\bibnamefont {Matsuda}}, \bibinfo {author} {\bibfnamefont {M.}~\bibnamefont {Holzmann}},\ and\ \bibinfo {author} {\bibfnamefont {D.~M.}\ \bibnamefont {Ceperley}},\ }\bibfield  {title} {\bibinfo {title} {{Quantum Monte Carlo Compton profiles of solid and liquid lithium}},\ }\href {https://doi.org/10.1103/PhysRevB.101.165125} {\bibfield  {journal} {\bibinfo  {journal} {Phys. Rev. B}\ }\textbf {\bibinfo {volume} {101}},\ \bibinfo {pages} {165125} (\bibinfo {year} {2020})}\BibitemShut {NoStop}%
\bibitem [{\citenamefont {Weiss}(1972)}]{Weiss_1972}%
  \BibitemOpen
  \bibfield  {author} {\bibinfo {author} {\bibfnamefont {R.~J.}\ \bibnamefont {Weiss}},\ }\bibfield  {title} {\bibinfo {title} {{Electron momentum distribution in silicon}},\ }\href {https://doi.org/10.1080/14786437208221026} {\bibfield  {journal} {\bibinfo  {journal} {The Philosophical Magazine: A Journal of Theoretical Experimental and Applied Physics}\ }\textbf {\bibinfo {volume} {26}},\ \bibinfo {pages} {153} (\bibinfo {year} {1972})}\BibitemShut {NoStop}%
\bibitem [{\citenamefont {Schülke}(1974)}]{shulke_1974}%
  \BibitemOpen
  \bibfield  {author} {\bibinfo {author} {\bibfnamefont {W.}~\bibnamefont {Schülke}},\ }\bibfield  {title} {\bibinfo {title} {{Reconstructed Three-Dimensional Valence Electron Momentum Density of Si from Compton Profile Measurements}},\ }\href {https://doi.org/https://doi.org/10.1002/pssb.2220620214} {\bibfield  {journal} {\bibinfo  {journal} {physica status solidi (b)}\ }\textbf {\bibinfo {volume} {62}},\ \bibinfo {pages} {453} (\bibinfo {year} {1974})},\ \Eprint {https://arxiv.org/abs/https://onlinelibrary.wiley.com/doi/pdf/10.1002/pssb.2220620214} {https://onlinelibrary.wiley.com/doi/pdf/10.1002/pssb.2220620214} \BibitemShut {NoStop}%
\bibitem [{\citenamefont {Seth}\ and\ \citenamefont {Ellis}(1977)}]{Seth_1977}%
  \BibitemOpen
  \bibfield  {author} {\bibinfo {author} {\bibfnamefont {A.}~\bibnamefont {Seth}}\ and\ \bibinfo {author} {\bibfnamefont {D.~E.}\ \bibnamefont {Ellis}},\ }\bibfield  {title} {\bibinfo {title} {{Momentum densities and Compton profiles of diamond, silicon and silicon carbide}},\ }\href {https://doi.org/10.1088/0022-3719/10/2/006} {\bibfield  {journal} {\bibinfo  {journal} {Journal of Physics C: Solid State Physics}\ }\textbf {\bibinfo {volume} {10}},\ \bibinfo {pages} {181} (\bibinfo {year} {1977})}\BibitemShut {NoStop}%
\bibitem [{\citenamefont {Pattison}\ and\ \citenamefont {Schneider}(1978)}]{Pattison_1978}%
  \BibitemOpen
  \bibfield  {author} {\bibinfo {author} {\bibfnamefont {P.}~\bibnamefont {Pattison}}\ and\ \bibinfo {author} {\bibfnamefont {J.}~\bibnamefont {Schneider}},\ }\bibfield  {title} {\bibinfo {title} {{Structural information from directional Compton profile measurements on Si and Ge}},\ }\href {https://doi.org/https://doi.org/10.1016/0038-1098(78)90494-5} {\bibfield  {journal} {\bibinfo  {journal} {Solid State Communications}\ }\textbf {\bibinfo {volume} {28}},\ \bibinfo {pages} {581} (\bibinfo {year} {1978})}\BibitemShut {NoStop}%
\bibitem [{\citenamefont {Pattison}\ \emph {et~al.}(1981)\citenamefont {Pattison}, \citenamefont {Hansen},\ and\ \citenamefont {Schneider}}]{Pattison_1981}%
  \BibitemOpen
  \bibfield  {author} {\bibinfo {author} {\bibfnamefont {P.}~\bibnamefont {Pattison}}, \bibinfo {author} {\bibfnamefont {N.}~\bibnamefont {Hansen}},\ and\ \bibinfo {author} {\bibfnamefont {J.}~\bibnamefont {Schneider}},\ }\bibfield  {title} {\bibinfo {title} {{Identifying the bonding in diamond and silicon using Compton scattering experiments}},\ }\href {https://doi.org/https://doi.org/10.1016/0301-0104(81)85166-X} {\bibfield  {journal} {\bibinfo  {journal} {Chemical Physics}\ }\textbf {\bibinfo {volume} {59}},\ \bibinfo {pages} {231} (\bibinfo {year} {1981})}\BibitemShut {NoStop}%
\bibitem [{\citenamefont {Hansen}\ \emph {et~al.}(1987)\citenamefont {Hansen}, \citenamefont {Pattison},\ and\ \citenamefont {Schneider}}]{hansen_1987}%
  \BibitemOpen
  \bibfield  {author} {\bibinfo {author} {\bibfnamefont {N.~K.}\ \bibnamefont {Hansen}}, \bibinfo {author} {\bibfnamefont {P.}~\bibnamefont {Pattison}},\ and\ \bibinfo {author} {\bibfnamefont {J.~R.}\ \bibnamefont {Schneider}},\ }\bibfield  {title} {\bibinfo {title} {{Analysis of the 3-dimensional electron distribution in silicon using directional Compton profile measurements}},\ }\href {https://doi.org/https://doi.org/10.1007/BF01305420} {\bibfield  {journal} {\bibinfo  {journal} {Zeitschrift f{\"u}r Physik B Condensed Matter}\ }\textbf {\bibinfo {volume} {66}},\ \bibinfo {pages} {305} (\bibinfo {year} {1987})}\BibitemShut {NoStop}%
\bibitem [{\citenamefont {Nara}\ \emph {et~al.}(1979)\citenamefont {Nara}, \citenamefont {Sindo},\ and\ \citenamefont {Kobayasi}}]{Nara_1979}%
  \BibitemOpen
  \bibfield  {author} {\bibinfo {author} {\bibfnamefont {H.}~\bibnamefont {Nara}}, \bibinfo {author} {\bibfnamefont {K.}~\bibnamefont {Sindo}},\ and\ \bibinfo {author} {\bibfnamefont {T.}~\bibnamefont {Kobayasi}},\ }\bibfield  {title} {\bibinfo {title} {{Pseudopotential Approach to Anisotropies of Compton-Profiles of Si and Ge}},\ }\href {https://doi.org/10.1143/JPSJ.46.77} {\bibfield  {journal} {\bibinfo  {journal} {Journal of the Physical Society of Japan}\ }\textbf {\bibinfo {volume} {46}},\ \bibinfo {pages} {77} (\bibinfo {year} {1979})}\BibitemShut {NoStop}%
\bibitem [{\citenamefont {Nara}\ \emph {et~al.}(1984)\citenamefont {Nara}, \citenamefont {Kobayasi},\ and\ \citenamefont {Shindo}}]{Nara_1984}%
  \BibitemOpen
  \bibfield  {author} {\bibinfo {author} {\bibfnamefont {H.}~\bibnamefont {Nara}}, \bibinfo {author} {\bibfnamefont {T.}~\bibnamefont {Kobayasi}},\ and\ \bibinfo {author} {\bibfnamefont {K.}~\bibnamefont {Shindo}},\ }\bibfield  {title} {\bibinfo {title} {{Anisotropies of Compton profiles of tetrahedrally bonded semiconductors}},\ }\href {https://doi.org/10.1088/0022-3719/17/22/015} {\bibfield  {journal} {\bibinfo  {journal} {Journal of Physics C: Solid State Physics}\ }\textbf {\bibinfo {volume} {17}},\ \bibinfo {pages} {3967} (\bibinfo {year} {1984})}\BibitemShut {NoStop}%
\bibitem [{\citenamefont {Sakai}\ \emph {et~al.}(1989)\citenamefont {Sakai}, \citenamefont {Shiotani}, \citenamefont {Itoh}, \citenamefont {Mao}, \citenamefont {Ito}, \citenamefont {Kawata}, \citenamefont {Amemiya},\ and\ \citenamefont {Ando}}]{Sakai_1989}%
  \BibitemOpen
  \bibfield  {author} {\bibinfo {author} {\bibfnamefont {N.}~\bibnamefont {Sakai}}, \bibinfo {author} {\bibfnamefont {N.}~\bibnamefont {Shiotani}}, \bibinfo {author} {\bibfnamefont {F.}~\bibnamefont {Itoh}}, \bibinfo {author} {\bibfnamefont {O.}~\bibnamefont {Mao}}, \bibinfo {author} {\bibfnamefont {M.}~\bibnamefont {Ito}}, \bibinfo {author} {\bibfnamefont {H.}~\bibnamefont {Kawata}}, \bibinfo {author} {\bibfnamefont {Y.}~\bibnamefont {Amemiya}},\ and\ \bibinfo {author} {\bibfnamefont {M.}~\bibnamefont {Ando}},\ }\bibfield  {title} {\bibinfo {title} {{High-Resolution Compton Profile of Si Using 29.5 keV Synchrotron-Radiation X-Rays}},\ }\href {https://doi.org/10.1143/JPSJ.58.3270} {\bibfield  {journal} {\bibinfo  {journal} {Journal of the Physical Society of Japan}\ }\textbf {\bibinfo {volume} {58}},\ \bibinfo {pages} {3270} (\bibinfo {year} {1989})}\BibitemShut {NoStop}%
\bibitem [{\citenamefont {Kubo}\ \emph {et~al.}(1997)\citenamefont {Kubo}, \citenamefont {Sakurai}, \citenamefont {Tanaka}, \citenamefont {Nakamura}, \citenamefont {Kawata},\ and\ \citenamefont {Shiotani}}]{Kubo_1997}%
  \BibitemOpen
  \bibfield  {author} {\bibinfo {author} {\bibfnamefont {Y.}~\bibnamefont {Kubo}}, \bibinfo {author} {\bibfnamefont {Y.}~\bibnamefont {Sakurai}}, \bibinfo {author} {\bibfnamefont {Y.}~\bibnamefont {Tanaka}}, \bibinfo {author} {\bibfnamefont {T.}~\bibnamefont {Nakamura}}, \bibinfo {author} {\bibfnamefont {H.}~\bibnamefont {Kawata}},\ and\ \bibinfo {author} {\bibfnamefont {N.}~\bibnamefont {Shiotani}},\ }\bibfield  {title} {\bibinfo {title} {{Effects of Self-Interaction Correction on Compton Profiles of Diamond and Silicon}},\ }\href {https://doi.org/10.1143/JPSJ.66.2777} {\bibfield  {journal} {\bibinfo  {journal} {Journal of the Physical Society of Japan}\ }\textbf {\bibinfo {volume} {66}},\ \bibinfo {pages} {2777} (\bibinfo {year} {1997})}\BibitemShut {NoStop}%
\bibitem [{\citenamefont {Delaney}\ \emph {et~al.}(1998)\citenamefont {Delaney}, \citenamefont {Kr\'alik},\ and\ \citenamefont {Louie}}]{Delaney_1998}%
  \BibitemOpen
  \bibfield  {author} {\bibinfo {author} {\bibfnamefont {P.}~\bibnamefont {Delaney}}, \bibinfo {author} {\bibfnamefont {B.}~\bibnamefont {Kr\'alik}},\ and\ \bibinfo {author} {\bibfnamefont {S.~G.}\ \bibnamefont {Louie}},\ }\bibfield  {title} {\bibinfo {title} {{Compton profiles of Si: Pseudopotential calculation and reconstruction effects}},\ }\href {https://doi.org/10.1103/PhysRevB.58.4320} {\bibfield  {journal} {\bibinfo  {journal} {Phys. Rev. B}\ }\textbf {\bibinfo {volume} {58}},\ \bibinfo {pages} {4320} (\bibinfo {year} {1998})}\BibitemShut {NoStop}%
\bibitem [{\citenamefont {Tse}\ \emph {et~al.}(2005)\citenamefont {Tse}, \citenamefont {Klug}, \citenamefont {Jiang}, \citenamefont {Sternemann}, \citenamefont {Volmer}, \citenamefont {Huotari}, \citenamefont {Hiraoka}, \citenamefont {Honkimäki},\ and\ \citenamefont {Hämäläinen}}]{Tse_2005}%
  \BibitemOpen
  \bibfield  {author} {\bibinfo {author} {\bibfnamefont {J.~S.}\ \bibnamefont {Tse}}, \bibinfo {author} {\bibfnamefont {D.~D.}\ \bibnamefont {Klug}}, \bibinfo {author} {\bibfnamefont {D.~T.}\ \bibnamefont {Jiang}}, \bibinfo {author} {\bibfnamefont {C.}~\bibnamefont {Sternemann}}, \bibinfo {author} {\bibfnamefont {M.}~\bibnamefont {Volmer}}, \bibinfo {author} {\bibfnamefont {S.}~\bibnamefont {Huotari}}, \bibinfo {author} {\bibfnamefont {N.}~\bibnamefont {Hiraoka}}, \bibinfo {author} {\bibfnamefont {V.}~\bibnamefont {Honkimäki}},\ and\ \bibinfo {author} {\bibfnamefont {K.}~\bibnamefont {Hämäläinen}},\ }\bibfield  {title} {\bibinfo {title} {{Compton scattering of elemental silicon at high pressure}},\ }\href {https://doi.org/10.1063/1.2126125} {\bibfield  {journal} {\bibinfo  {journal} {Applied Physics Letters}\ }\textbf {\bibinfo {volume} {87}},\ \bibinfo {pages} {191905} (\bibinfo {year} {2005})}\BibitemShut {NoStop}%
\bibitem [{\citenamefont {Okada}\ \emph {et~al.}(2012)\citenamefont {Okada}, \citenamefont {Sit}, \citenamefont {Watanabe}, \citenamefont {Wang}, \citenamefont {Barbiellini}, \citenamefont {Ishikawa}, \citenamefont {Itou}, \citenamefont {Sakurai}, \citenamefont {Bansil}, \citenamefont {Ishikawa}, \citenamefont {Hamaishi}, \citenamefont {Masaki}, \citenamefont {Paradis}, \citenamefont {Kimura}, \citenamefont {Ishikawa},\ and\ \citenamefont {Nanao}}]{Okada_2012}%
  \BibitemOpen
  \bibfield  {author} {\bibinfo {author} {\bibfnamefont {J.~T.}\ \bibnamefont {Okada}}, \bibinfo {author} {\bibfnamefont {P.~H.-L.}\ \bibnamefont {Sit}}, \bibinfo {author} {\bibfnamefont {Y.}~\bibnamefont {Watanabe}}, \bibinfo {author} {\bibfnamefont {Y.~J.}\ \bibnamefont {Wang}}, \bibinfo {author} {\bibfnamefont {B.}~\bibnamefont {Barbiellini}}, \bibinfo {author} {\bibfnamefont {T.}~\bibnamefont {Ishikawa}}, \bibinfo {author} {\bibfnamefont {M.}~\bibnamefont {Itou}}, \bibinfo {author} {\bibfnamefont {Y.}~\bibnamefont {Sakurai}}, \bibinfo {author} {\bibfnamefont {A.}~\bibnamefont {Bansil}}, \bibinfo {author} {\bibfnamefont {R.}~\bibnamefont {Ishikawa}}, \bibinfo {author} {\bibfnamefont {M.}~\bibnamefont {Hamaishi}}, \bibinfo {author} {\bibfnamefont {T.}~\bibnamefont {Masaki}}, \bibinfo {author} {\bibfnamefont {P.-F.}\ \bibnamefont {Paradis}}, \bibinfo {author} {\bibfnamefont {K.}~\bibnamefont {Kimura}}, \bibinfo {author} {\bibfnamefont {T.}~\bibnamefont {Ishikawa}},\ and\ \bibinfo {author} {\bibfnamefont
  {S.}~\bibnamefont {Nanao}},\ }\bibfield  {title} {\bibinfo {title} {{Persistence of Covalent Bonding in Liquid Silicon Probed by Inelastic X-Ray Scattering}},\ }\href {https://doi.org/10.1103/PhysRevLett.108.067402} {\bibfield  {journal} {\bibinfo  {journal} {Phys. Rev. Lett.}\ }\textbf {\bibinfo {volume} {108}},\ \bibinfo {pages} {067402} (\bibinfo {year} {2012})}\BibitemShut {NoStop}%
\bibitem [{\citenamefont {Matsuda}\ \emph {et~al.}(2013)\citenamefont {Matsuda}, \citenamefont {Nagao}, \citenamefont {Kajihara}, \citenamefont {Inui}, \citenamefont {Tamura}, \citenamefont {Nakamura}, \citenamefont {Kimura}, \citenamefont {Yao}, \citenamefont {Itou}, \citenamefont {Sakurai},\ and\ \citenamefont {Hiraoka}}]{Matsuda_2013}%
  \BibitemOpen
  \bibfield  {author} {\bibinfo {author} {\bibfnamefont {K.}~\bibnamefont {Matsuda}}, \bibinfo {author} {\bibfnamefont {T.}~\bibnamefont {Nagao}}, \bibinfo {author} {\bibfnamefont {Y.}~\bibnamefont {Kajihara}}, \bibinfo {author} {\bibfnamefont {M.}~\bibnamefont {Inui}}, \bibinfo {author} {\bibfnamefont {K.}~\bibnamefont {Tamura}}, \bibinfo {author} {\bibfnamefont {J.}~\bibnamefont {Nakamura}}, \bibinfo {author} {\bibfnamefont {K.}~\bibnamefont {Kimura}}, \bibinfo {author} {\bibfnamefont {M.}~\bibnamefont {Yao}}, \bibinfo {author} {\bibfnamefont {M.}~\bibnamefont {Itou}}, \bibinfo {author} {\bibfnamefont {Y.}~\bibnamefont {Sakurai}},\ and\ \bibinfo {author} {\bibfnamefont {N.}~\bibnamefont {Hiraoka}},\ }\bibfield  {title} {\bibinfo {title} {{Electron momentum density in liquid silicon}},\ }\href {https://doi.org/10.1103/PhysRevB.88.115125} {\bibfield  {journal} {\bibinfo  {journal} {Phys. Rev. B}\ }\textbf {\bibinfo {volume} {88}},\ \bibinfo {pages} {115125} (\bibinfo {year} {2013})}\BibitemShut {NoStop}%
\bibitem [{\citenamefont {Klevak}\ \emph {et~al.}(2016)\citenamefont {Klevak}, \citenamefont {Vila}, \citenamefont {Kas}, \citenamefont {Rehr},\ and\ \citenamefont {Seidler}}]{Klevak_2016}%
  \BibitemOpen
  \bibfield  {author} {\bibinfo {author} {\bibfnamefont {E.}~\bibnamefont {Klevak}}, \bibinfo {author} {\bibfnamefont {F.~D.}\ \bibnamefont {Vila}}, \bibinfo {author} {\bibfnamefont {J.~J.}\ \bibnamefont {Kas}}, \bibinfo {author} {\bibfnamefont {J.~J.}\ \bibnamefont {Rehr}},\ and\ \bibinfo {author} {\bibfnamefont {G.~T.}\ \bibnamefont {Seidler}},\ }\bibfield  {title} {\bibinfo {title} {{Finite-temperature calculations of the Compton profile of Be, Li, and Si}},\ }\href {https://doi.org/10.1103/PhysRevB.94.214201} {\bibfield  {journal} {\bibinfo  {journal} {Phys. Rev. B}\ }\textbf {\bibinfo {volume} {94}},\ \bibinfo {pages} {214201} (\bibinfo {year} {2016})}\BibitemShut {NoStop}%
\bibitem [{\citenamefont {Aguiar}\ and\ \citenamefont {{Di~Rocco}}(2026)}]{Aguiar_2026}%
  \BibitemOpen
  \bibfield  {author} {\bibinfo {author} {\bibfnamefont {J.~C.}\ \bibnamefont {Aguiar}}\ and\ \bibinfo {author} {\bibfnamefont {H.~O.}\ \bibnamefont {{Di~Rocco}}},\ }\bibfield  {title} {\bibinfo {title} {{Spherically averaged Hartree-Fock orbitals in an APW-like framework: Comparison with Compton profile experiments}},\ }\href {https://doi.org/https://doi.org/10.1016/j.physleta.2026.131387} {\bibfield  {journal} {\bibinfo  {journal} {Physics Letters A}\ }\textbf {\bibinfo {volume} {575}},\ \bibinfo {pages} {131387} (\bibinfo {year} {2026})}\BibitemShut {NoStop}%
\bibitem [{\citenamefont {van Schilfgaarde}\ \emph {et~al.}(2006)\citenamefont {van Schilfgaarde}, \citenamefont {Kotani},\ and\ \citenamefont {Faleev}}]{van_schilfgaarde_adequacy_2006}%
  \BibitemOpen
  \bibfield  {author} {\bibinfo {author} {\bibfnamefont {M.}~\bibnamefont {van Schilfgaarde}}, \bibinfo {author} {\bibfnamefont {T.}~\bibnamefont {Kotani}},\ and\ \bibinfo {author} {\bibfnamefont {S.~V.}\ \bibnamefont {Faleev}},\ }\bibfield  {title} {\bibinfo {title} {Adequacy of approximations in {GW} theory},\ }\href {https://doi.org/10.1103/PhysRevB.74.245125} {\bibfield  {journal} {\bibinfo  {journal} {Phys. Rev. B}\ }\textbf {\bibinfo {volume} {74}},\ \bibinfo {pages} {245125} (\bibinfo {year} {2006})}\BibitemShut {NoStop}%
\bibitem [{\citenamefont {Faleev}\ \emph {et~al.}(2006)\citenamefont {Faleev}, \citenamefont {van Schilfgaarde}, \citenamefont {Kotani}, \citenamefont {L\'eonard},\ and\ \citenamefont {Desjarlais}}]{Faleev_2006}%
  \BibitemOpen
  \bibfield  {author} {\bibinfo {author} {\bibfnamefont {S.~V.}\ \bibnamefont {Faleev}}, \bibinfo {author} {\bibfnamefont {M.}~\bibnamefont {van Schilfgaarde}}, \bibinfo {author} {\bibfnamefont {T.}~\bibnamefont {Kotani}}, \bibinfo {author} {\bibfnamefont {F.}~\bibnamefont {L\'eonard}},\ and\ \bibinfo {author} {\bibfnamefont {M.~P.}\ \bibnamefont {Desjarlais}},\ }\bibfield  {title} {\bibinfo {title} {{Finite-temperature quasiparticle self-consistent $\mathit{GW}$ approximation}},\ }\href {https://doi.org/10.1103/PhysRevB.74.033101} {\bibfield  {journal} {\bibinfo  {journal} {Phys. Rev. B}\ }\textbf {\bibinfo {volume} {74}},\ \bibinfo {pages} {033101} (\bibinfo {year} {2006})}\BibitemShut {NoStop}%
\bibitem [{\citenamefont {Kotani}\ \emph {et~al.}(2007{\natexlab{b}})\citenamefont {Kotani}, \citenamefont {van Schilfgaarde},\ and\ \citenamefont {Faleev}}]{kotani_quasiparticle_2007}%
  \BibitemOpen
  \bibfield  {author} {\bibinfo {author} {\bibfnamefont {T.}~\bibnamefont {Kotani}}, \bibinfo {author} {\bibfnamefont {M.}~\bibnamefont {van Schilfgaarde}},\ and\ \bibinfo {author} {\bibfnamefont {S.~V.}\ \bibnamefont {Faleev}},\ }\bibfield  {title} {\bibinfo {title} {Quasiparticle self-consistent {GW} method: {A} basis for the independent-particle approximation},\ }\href {https://doi.org/10.1103/PhysRevB.76.165106} {\bibfield  {journal} {\bibinfo  {journal} {Phys. Rev. B}\ }\textbf {\bibinfo {volume} {76}},\ \bibinfo {pages} {165106} (\bibinfo {year} {2007}{\natexlab{b}})}\BibitemShut {NoStop}%
\bibitem [{\citenamefont {Grumet}\ \emph {et~al.}(2018)\citenamefont {Grumet}, \citenamefont {Liu}, \citenamefont {Kaltak}, \citenamefont {Klime{\v{s}}},\ and\ \citenamefont {Kresse}}]{Grumet_2018}%
  \BibitemOpen
  \bibfield  {author} {\bibinfo {author} {\bibfnamefont {M.}~\bibnamefont {Grumet}}, \bibinfo {author} {\bibfnamefont {P.}~\bibnamefont {Liu}}, \bibinfo {author} {\bibfnamefont {M.}~\bibnamefont {Kaltak}}, \bibinfo {author} {\bibfnamefont {J.}~\bibnamefont {Klime{\v{s}}}},\ and\ \bibinfo {author} {\bibfnamefont {G.}~\bibnamefont {Kresse}},\ }\bibfield  {title} {\bibinfo {title} {{Beyond the quasiparticle approximation: Fully self-consistent $GW$ calculations}},\ }\href {https://doi.org/10.1103/PhysRevB.98.155143} {\bibfield  {journal} {\bibinfo  {journal} {Phys. Rev. B}\ }\textbf {\bibinfo {volume} {98}},\ \bibinfo {pages} {155143} (\bibinfo {year} {2018})}\BibitemShut {NoStop}%
\bibitem [{\citenamefont {Salas-Illanes}\ \emph {et~al.}(2022)\citenamefont {Salas-Illanes}, \citenamefont {Nabok},\ and\ \citenamefont {Draxl}}]{Salas-Illanes_2022}%
  \BibitemOpen
  \bibfield  {author} {\bibinfo {author} {\bibfnamefont {N.}~\bibnamefont {Salas-Illanes}}, \bibinfo {author} {\bibfnamefont {D.}~\bibnamefont {Nabok}},\ and\ \bibinfo {author} {\bibfnamefont {C.}~\bibnamefont {Draxl}},\ }\bibfield  {title} {\bibinfo {title} {{Electronic structure of representative band-gap materials by all-electron quasiparticle self-consistent $GW$ calculations}},\ }\href {https://doi.org/10.1103/PhysRevB.106.045143} {\bibfield  {journal} {\bibinfo  {journal} {Phys. Rev. B}\ }\textbf {\bibinfo {volume} {106}},\ \bibinfo {pages} {045143} (\bibinfo {year} {2022})}\BibitemShut {NoStop}%
\bibitem [{\citenamefont {Harris-Lee}\ \emph {et~al.}(2021)\citenamefont {Harris-Lee}, \citenamefont {James},\ and\ \citenamefont {Dugdale}}]{harris-lee_sensitivity_2021}%
  \BibitemOpen
  \bibfield  {author} {\bibinfo {author} {\bibfnamefont {E.~I.}\ \bibnamefont {Harris-Lee}}, \bibinfo {author} {\bibfnamefont {A.~D.~N.}\ \bibnamefont {James}},\ and\ \bibinfo {author} {\bibfnamefont {S.~B.}\ \bibnamefont {Dugdale}},\ }\bibfield  {title} {\bibinfo {title} {Sensitivity of the {Fermi} surface to the treatment of exchange and correlation},\ }\href {https://doi.org/10.1103/PhysRevB.103.235144} {\bibfield  {journal} {\bibinfo  {journal} {Phys. Rev. B}\ }\textbf {\bibinfo {volume} {103}},\ \bibinfo {pages} {235144} (\bibinfo {year} {2021})}\BibitemShut {NoStop}%
\bibitem [{\citenamefont {Tanaka}\ \emph {et~al.}(2000)\citenamefont {Tanaka}, \citenamefont {Chen}, \citenamefont {Bellin}, \citenamefont {Loupias}, \citenamefont {Fretwell}, \citenamefont {Rodrigues-Gonzalez}, \citenamefont {Alam}, \citenamefont {Dugdale}, \citenamefont {Manuel}, \citenamefont {Shukla}, \citenamefont {Buslaps}, \citenamefont {Suortti},\ and\ \citenamefont {Shiotani}}]{tanaka_study_2000}%
  \BibitemOpen
  \bibfield  {author} {\bibinfo {author} {\bibfnamefont {Y.}~\bibnamefont {Tanaka}}, \bibinfo {author} {\bibfnamefont {K.~J.}\ \bibnamefont {Chen}}, \bibinfo {author} {\bibfnamefont {C.}~\bibnamefont {Bellin}}, \bibinfo {author} {\bibfnamefont {G.}~\bibnamefont {Loupias}}, \bibinfo {author} {\bibfnamefont {H.~M.}\ \bibnamefont {Fretwell}}, \bibinfo {author} {\bibfnamefont {A.}~\bibnamefont {Rodrigues-Gonzalez}}, \bibinfo {author} {\bibfnamefont {M.~A.}\ \bibnamefont {Alam}}, \bibinfo {author} {\bibfnamefont {S.~B.}\ \bibnamefont {Dugdale}}, \bibinfo {author} {\bibfnamefont {A.~A.}\ \bibnamefont {Manuel}}, \bibinfo {author} {\bibfnamefont {A.}~\bibnamefont {Shukla}}, \bibinfo {author} {\bibfnamefont {T.}~\bibnamefont {Buslaps}}, \bibinfo {author} {\bibfnamefont {P.}~\bibnamefont {Suortti}},\ and\ \bibinfo {author} {\bibfnamefont {N.}~\bibnamefont {Shiotani}},\ }\bibfield  {title} {\bibinfo {title} {A study on the {Fermi} surface of {Cr} by high-resolution {Compton} scattering},\ }\href
  {https://doi.org/10.1016/S0022-3697(99)00318-2} {\bibfield  {journal} {\bibinfo  {journal} {Journal of Physics and Chemistry of Solids}\ }\textbf {\bibinfo {volume} {61}},\ \bibinfo {pages} {365} (\bibinfo {year} {2000})}\BibitemShut {NoStop}%
\bibitem [{\citenamefont {Fawcett}(1988)}]{RevModPhys.60.209}%
  \BibitemOpen
  \bibfield  {author} {\bibinfo {author} {\bibfnamefont {E.}~\bibnamefont {Fawcett}},\ }\bibfield  {title} {\bibinfo {title} {Spin-density-wave antiferromagnetism in chromium},\ }\href {https://doi.org/10.1103/RevModPhys.60.209} {\bibfield  {journal} {\bibinfo  {journal} {Rev. Mod. Phys.}\ }\textbf {\bibinfo {volume} {60}},\ \bibinfo {pages} {209} (\bibinfo {year} {1988})}\BibitemShut {NoStop}%
\bibitem [{\citenamefont {Fretwell}\ \emph {et~al.}(1995)\citenamefont {Fretwell}, \citenamefont {Dugdale}, \citenamefont {Alam}, \citenamefont {Biasini}, \citenamefont {Hoffmann},\ and\ \citenamefont {Manuel}}]{Fretwell_1995}%
  \BibitemOpen
  \bibfield  {author} {\bibinfo {author} {\bibfnamefont {H.~M.}\ \bibnamefont {Fretwell}}, \bibinfo {author} {\bibfnamefont {S.~B.}\ \bibnamefont {Dugdale}}, \bibinfo {author} {\bibfnamefont {M.~A.}\ \bibnamefont {Alam}}, \bibinfo {author} {\bibfnamefont {M.}~\bibnamefont {Biasini}}, \bibinfo {author} {\bibfnamefont {L.}~\bibnamefont {Hoffmann}},\ and\ \bibinfo {author} {\bibfnamefont {A.~A.}\ \bibnamefont {Manuel}},\ }\bibfield  {title} {\bibinfo {title} {{Reconstruction of 3D Electron-Positron Momentum Densities from 2D Projections: Role of Maximum-Entropy Deconvolution Prior to Reconstruction}},\ }\href {https://doi.org/10.1209/0295-5075/32/9/012} {\bibfield  {journal} {\bibinfo  {journal} {Europhysics Letters}\ }\textbf {\bibinfo {volume} {32}},\ \bibinfo {pages} {771} (\bibinfo {year} {1995})}\BibitemShut {NoStop}%
\bibitem [{\citenamefont {Matsumoto}\ and\ \citenamefont {Wakoh}(1986)}]{matsumoto_86}%
  \BibitemOpen
  \bibfield  {author} {\bibinfo {author} {\bibfnamefont {M.}~\bibnamefont {Matsumoto}}\ and\ \bibinfo {author} {\bibfnamefont {S.}~\bibnamefont {Wakoh}},\ }\bibfield  {title} {\bibinfo {title} {{Two-Dimensional Angular Correlation Distributions of Positron Annihilation Radiation in Vanadium and Chromium}},\ }\href {https://doi.org/10.1143/JPSJ.55.3948} {\bibfield  {journal} {\bibinfo  {journal} {Journal of the Physical Society of Japan}\ }\textbf {\bibinfo {volume} {55}},\ \bibinfo {pages} {3948} (\bibinfo {year} {1986})}\BibitemShut {NoStop}%
\bibitem [{\citenamefont {Dugdale}\ \emph {et~al.}(1998)\citenamefont {Dugdale}, \citenamefont {Fretwell}, \citenamefont {Hedley}, \citenamefont {Alam}, \citenamefont {Jarlborg}, \citenamefont {Santi}, \citenamefont {Singru}, \citenamefont {Sundararajan},\ and\ \citenamefont {Cooper}}]{dugdale_fermiology_1998}%
  \BibitemOpen
  \bibfield  {author} {\bibinfo {author} {\bibfnamefont {S.~B.}\ \bibnamefont {Dugdale}}, \bibinfo {author} {\bibfnamefont {H.~M.}\ \bibnamefont {Fretwell}}, \bibinfo {author} {\bibfnamefont {D.~C.~R.}\ \bibnamefont {Hedley}}, \bibinfo {author} {\bibfnamefont {M.~A.}\ \bibnamefont {Alam}}, \bibinfo {author} {\bibfnamefont {T.}~\bibnamefont {Jarlborg}}, \bibinfo {author} {\bibfnamefont {G.}~\bibnamefont {Santi}}, \bibinfo {author} {\bibfnamefont {R.~M.}\ \bibnamefont {Singru}}, \bibinfo {author} {\bibfnamefont {V.}~\bibnamefont {Sundararajan}},\ and\ \bibinfo {author} {\bibfnamefont {M.~J.}\ \bibnamefont {Cooper}},\ }\bibfield  {title} {\bibinfo {title} {Fermiology of {Cr} and {Mo}},\ }\href {https://doi.org/10.1088/0953-8984/10/46/003} {\bibfield  {journal} {\bibinfo  {journal} {J. Phys.: Condens. Matter}\ }\textbf {\bibinfo {volume} {10}},\ \bibinfo {pages} {10367} (\bibinfo {year} {1998})}\BibitemShut {NoStop}%
\bibitem [{\citenamefont {Hughes}\ \emph {et~al.}(2004)\citenamefont {Hughes}, \citenamefont {Dugdale}, \citenamefont {Major}, \citenamefont {Alam}, \citenamefont {Jarlborg}, \citenamefont {Bruno},\ and\ \citenamefont {Ginatempo}}]{PhysRevB.69.174406}%
  \BibitemOpen
  \bibfield  {author} {\bibinfo {author} {\bibfnamefont {R.~J.}\ \bibnamefont {Hughes}}, \bibinfo {author} {\bibfnamefont {S.~B.}\ \bibnamefont {Dugdale}}, \bibinfo {author} {\bibfnamefont {Z.}~\bibnamefont {Major}}, \bibinfo {author} {\bibfnamefont {M.~A.}\ \bibnamefont {Alam}}, \bibinfo {author} {\bibfnamefont {T.}~\bibnamefont {Jarlborg}}, \bibinfo {author} {\bibfnamefont {E.}~\bibnamefont {Bruno}},\ and\ \bibinfo {author} {\bibfnamefont {B.}~\bibnamefont {Ginatempo}},\ }\bibfield  {title} {\bibinfo {title} {{Evolution of the Fermi surface and the oscillatory exchange coupling across Cr and Cr-based alloys}},\ }\href {https://doi.org/10.1103/PhysRevB.69.174406} {\bibfield  {journal} {\bibinfo  {journal} {Phys. Rev. B}\ }\textbf {\bibinfo {volume} {69}},\ \bibinfo {pages} {174406} (\bibinfo {year} {2004})}\BibitemShut {NoStop}%
\bibitem [{\citenamefont {Graebner}\ and\ \citenamefont {Marcus}(1968)}]{graebner_haas-van_1968}%
  \BibitemOpen
  \bibfield  {author} {\bibinfo {author} {\bibfnamefont {J.~E.}\ \bibnamefont {Graebner}}\ and\ \bibinfo {author} {\bibfnamefont {J.~A.}\ \bibnamefont {Marcus}},\ }\bibfield  {title} {\bibinfo {title} {de {Haas}-van {Alphen} {Effect} in {Antiferromagnetic} {Chromium}},\ }\href {https://doi.org/10.1103/PhysRev.175.659} {\bibfield  {journal} {\bibinfo  {journal} {Phys. Rev.}\ }\textbf {\bibinfo {volume} {175}},\ \bibinfo {pages} {659} (\bibinfo {year} {1968})}\BibitemShut {NoStop}%
\bibitem [{\citenamefont {Laurent}\ \emph {et~al.}(1981)\citenamefont {Laurent}, \citenamefont {Callaway}, \citenamefont {Fry},\ and\ \citenamefont {Brener}}]{laurent_band_1981}%
  \BibitemOpen
  \bibfield  {author} {\bibinfo {author} {\bibfnamefont {D.~G.}\ \bibnamefont {Laurent}}, \bibinfo {author} {\bibfnamefont {J.}~\bibnamefont {Callaway}}, \bibinfo {author} {\bibfnamefont {J.~L.}\ \bibnamefont {Fry}},\ and\ \bibinfo {author} {\bibfnamefont {N.~E.}\ \bibnamefont {Brener}},\ }\bibfield  {title} {\bibinfo {title} {Band structure, {Fermi} surface, {Compton} profile, and optical conductivity of paramagnetic chromium},\ }\href {https://doi.org/10.1103/PhysRevB.23.4977} {\bibfield  {journal} {\bibinfo  {journal} {Phys. Rev. B}\ }\textbf {\bibinfo {volume} {23}},\ \bibinfo {pages} {4977} (\bibinfo {year} {1981})}\BibitemShut {NoStop}%
\bibitem [{\citenamefont {K\"{u}bler}(1980)}]{kubler_spin-density_1980}%
  \BibitemOpen
  \bibfield  {author} {\bibinfo {author} {\bibfnamefont {J.}~\bibnamefont {K\"{u}bler}},\ }\bibfield  {title} {\bibinfo {title} {{Spin-density functional calculations for chromium}},\ }\href {https://doi.org/10.1016/0304-8853(80)90446-1} {\bibfield  {journal} {\bibinfo  {journal} {J. Magn. Magn. Mater.}\ }\textbf {\bibinfo {volume} {20}},\ \bibinfo {pages} {277} (\bibinfo {year} {1980})}\BibitemShut {NoStop}%
\bibitem [{\citenamefont {Sakai}\ \emph {et~al.}(1991)\citenamefont {Sakai}, \citenamefont {Ito}, \citenamefont {Kawata}, \citenamefont {Iwazumi}, \citenamefont {Ando}, \citenamefont {Shiotani}, \citenamefont {Itoh}, \citenamefont {Sakurai},\ and\ \citenamefont {Nanao}}]{sakai_application_1991}%
  \BibitemOpen
  \bibfield  {author} {\bibinfo {author} {\bibfnamefont {N.}~\bibnamefont {Sakai}}, \bibinfo {author} {\bibfnamefont {M.}~\bibnamefont {Ito}}, \bibinfo {author} {\bibfnamefont {H.}~\bibnamefont {Kawata}}, \bibinfo {author} {\bibfnamefont {T.}~\bibnamefont {Iwazumi}}, \bibinfo {author} {\bibfnamefont {M.}~\bibnamefont {Ando}}, \bibinfo {author} {\bibfnamefont {N.}~\bibnamefont {Shiotani}}, \bibinfo {author} {\bibfnamefont {F.}~\bibnamefont {Itoh}}, \bibinfo {author} {\bibfnamefont {Y.}~\bibnamefont {Sakurai}},\ and\ \bibinfo {author} {\bibfnamefont {S.}~\bibnamefont {Nanao}},\ }\bibfield  {title} {\bibinfo {title} {Application of circularly polarized x-rays to magnetic {Compton}-scattering experiments},\ }\href {https://doi.org/10.1016/0168-9002(91)90285-X} {\bibfield  {journal} {\bibinfo  {journal} {Nuclear Instruments and Methods in Physics Research Section A: Accelerators, Spectrometers, Detectors and Associated Equipment}\ }\textbf {\bibinfo {volume} {303}},\ \bibinfo {pages} {488} (\bibinfo {year}
  {1991})}\BibitemShut {NoStop}%
\bibitem [{\citenamefont {Timms}\ \emph {et~al.}(1990)\citenamefont {Timms}, \citenamefont {Brahmia}, \citenamefont {Cooper}, \citenamefont {Collins}, \citenamefont {Hamouda}, \citenamefont {Laundy}, \citenamefont {Kilbourne},\ and\ \citenamefont {Lager}}]{timms_spin_1990}%
  \BibitemOpen
  \bibfield  {author} {\bibinfo {author} {\bibfnamefont {D.~N.}\ \bibnamefont {Timms}}, \bibinfo {author} {\bibfnamefont {A.}~\bibnamefont {Brahmia}}, \bibinfo {author} {\bibfnamefont {M.~J.}\ \bibnamefont {Cooper}}, \bibinfo {author} {\bibfnamefont {S.~P.}\ \bibnamefont {Collins}}, \bibinfo {author} {\bibfnamefont {S.}~\bibnamefont {Hamouda}}, \bibinfo {author} {\bibfnamefont {D.}~\bibnamefont {Laundy}}, \bibinfo {author} {\bibfnamefont {C.}~\bibnamefont {Kilbourne}},\ and\ \bibinfo {author} {\bibfnamefont {M.-C.~S.}\ \bibnamefont {Lager}},\ }\bibfield  {title} {\bibinfo {title} {Spin dependent anisotropy in the momentum density of ferromagnetic nickel metal},\ }\href {https://doi.org/10.1088/0953-8984/2/14/028} {\bibfield  {journal} {\bibinfo  {journal} {J. Phys.: Condens. Matter}\ }\textbf {\bibinfo {volume} {2}},\ \bibinfo {pages} {3427} (\bibinfo {year} {1990})}\BibitemShut {NoStop}%
\bibitem [{\citenamefont {Billington}\ \emph {et~al.}(2020)\citenamefont {Billington}, \citenamefont {James}, \citenamefont {Harris-Lee}, \citenamefont {Lagos}, \citenamefont {O'Neill}, \citenamefont {Tsuda}, \citenamefont {Toyoki}, \citenamefont {Kotani}, \citenamefont {Nakamura}, \citenamefont {Bei}, \citenamefont {Mu}, \citenamefont {Samolyuk}, \citenamefont {Stocks}, \citenamefont {Duffy}, \citenamefont {Taylor}, \citenamefont {Giblin},\ and\ \citenamefont {Dugdale}}]{billington_2020}%
  \BibitemOpen
  \bibfield  {author} {\bibinfo {author} {\bibfnamefont {D.}~\bibnamefont {Billington}}, \bibinfo {author} {\bibfnamefont {A.~D.~N.}\ \bibnamefont {James}}, \bibinfo {author} {\bibfnamefont {E.~I.}\ \bibnamefont {Harris-Lee}}, \bibinfo {author} {\bibfnamefont {D.~A.}\ \bibnamefont {Lagos}}, \bibinfo {author} {\bibfnamefont {D.}~\bibnamefont {O'Neill}}, \bibinfo {author} {\bibfnamefont {N.}~\bibnamefont {Tsuda}}, \bibinfo {author} {\bibfnamefont {K.}~\bibnamefont {Toyoki}}, \bibinfo {author} {\bibfnamefont {Y.}~\bibnamefont {Kotani}}, \bibinfo {author} {\bibfnamefont {T.}~\bibnamefont {Nakamura}}, \bibinfo {author} {\bibfnamefont {H.}~\bibnamefont {Bei}}, \bibinfo {author} {\bibfnamefont {S.}~\bibnamefont {Mu}}, \bibinfo {author} {\bibfnamefont {G.~D.}\ \bibnamefont {Samolyuk}}, \bibinfo {author} {\bibfnamefont {G.~M.}\ \bibnamefont {Stocks}}, \bibinfo {author} {\bibfnamefont {J.~A.}\ \bibnamefont {Duffy}}, \bibinfo {author} {\bibfnamefont {J.~W.}\ \bibnamefont {Taylor}}, \bibinfo {author} {\bibfnamefont
  {S.~R.}\ \bibnamefont {Giblin}},\ and\ \bibinfo {author} {\bibfnamefont {S.~B.}\ \bibnamefont {Dugdale}},\ }\bibfield  {title} {\bibinfo {title} {Bulk and element-specific magnetism of medium-entropy and high-entropy {C}antor-{W}u alloys},\ }\href {https://doi.org/10.1103/PhysRevB.102.174405} {\bibfield  {journal} {\bibinfo  {journal} {Phys. Rev. B}\ }\textbf {\bibinfo {volume} {102}},\ \bibinfo {pages} {174405} (\bibinfo {year} {2020})}\BibitemShut {NoStop}%
\bibitem [{\citenamefont {Kubo}(2004)}]{kubo_electron_2004}%
  \BibitemOpen
  \bibfield  {author} {\bibinfo {author} {\bibfnamefont {Y.}~\bibnamefont {Kubo}},\ }\bibfield  {title} {\bibinfo {title} {Electron correlation effects on magnetic {Compton} profiles of nickel in the {GW} approximation},\ }\href {https://doi.org/10.1016/j.jpcs.2004.08.023} {\bibfield  {journal} {\bibinfo  {journal} {Journal of Physics and Chemistry of Solids}\ }\bibinfo {series} {Sagamore {XIV}: {Charge}, {Spin} and {Momentum} {Densities}},\ \textbf {\bibinfo {volume} {65}},\ \bibinfo {pages} {2077} (\bibinfo {year} {2004})}\BibitemShut {NoStop}%
\bibitem [{\citenamefont {Sponza}\ \emph {et~al.}(2017)\citenamefont {Sponza}, \citenamefont {Pisanti}, \citenamefont {Vishina}, \citenamefont {Pashov}, \citenamefont {Weber}, \citenamefont {van Schilfgaarde}, \citenamefont {Acharya}, \citenamefont {Vidal},\ and\ \citenamefont {Kotliar}}]{sponza_2017}%
  \BibitemOpen
  \bibfield  {author} {\bibinfo {author} {\bibfnamefont {L.}~\bibnamefont {Sponza}}, \bibinfo {author} {\bibfnamefont {P.}~\bibnamefont {Pisanti}}, \bibinfo {author} {\bibfnamefont {A.}~\bibnamefont {Vishina}}, \bibinfo {author} {\bibfnamefont {D.}~\bibnamefont {Pashov}}, \bibinfo {author} {\bibfnamefont {C.}~\bibnamefont {Weber}}, \bibinfo {author} {\bibfnamefont {M.}~\bibnamefont {van Schilfgaarde}}, \bibinfo {author} {\bibfnamefont {S.}~\bibnamefont {Acharya}}, \bibinfo {author} {\bibfnamefont {J.}~\bibnamefont {Vidal}},\ and\ \bibinfo {author} {\bibfnamefont {G.}~\bibnamefont {Kotliar}},\ }\bibfield  {title} {\bibinfo {title} {Self-energies in itinerant magnets: {A} focus on {Fe} and {Ni}},\ }\href {https://doi.org/10.1103/PhysRevB.95.041112} {\bibfield  {journal} {\bibinfo  {journal} {Phys. Rev. B}\ }\textbf {\bibinfo {volume} {95}},\ \bibinfo {pages} {041112} (\bibinfo {year} {2017})}\BibitemShut {NoStop}%
\end{thebibliography}
\end{document}